\RequirePackage{fix-cm}
\documentclass[twocolumn,epjc3]{svjour3}  
\smartqed  
\RequirePackage{graphicx}
\usepackage[T1,T2A]{fontenc}
\usepackage{xcolor}
\journalname{Eur. Phys. J. A}
\begin{document}

\title{Accessing tens-to-hundreds femtoseconds nuclear state lifetimes with low-energy binary heavy-ion reactions
}


\author{M. Ciema{\l}a\thanksref{e1,IFJPAN}
        \and
        S. Ziliani\thanksref{UNIMI,INFNMI}
        \and
		F. C. L. Crespi\thanksref{UNIMI,INFNMI}
		\and
		S. Leoni\thanksref{e2,UNIMI,INFNMI}
		\and 
		B. Fornal\thanksref{e3,IFJPAN} 
		\and \\
		A. Maj\thanksref{IFJPAN} 
		\and  
		P. Bednarczyk\thanksref{IFJPAN} 
		\and
		G. Benzoni\thanksref{INFNMI}
		\and
		A. Bracco\thanksref{UNIMI,INFNMI}
		\and 
		C. Boiano\thanksref{INFNMI} 
		\and \\
		S. Bottoni\thanksref{UNIMI,INFNMI}
		\and 
		S. Brambilla\thanksref{INFNMI} 
		\and 
		 M. Bast\thanksref{IKPCol}
		 \and
		 M.  Beckers\thanksref{IKPCol}
		 \and
		  T. Braunroth\thanksref{IKPCol}
		  \and \\
		F. Camera\thanksref{UNIMI,INFNMI}
		\and 
		N. Cieplicka-Ory\`nczak\thanksref{IFJPAN}
		\and
		E. Cl\'ement\thanksref{GANIL} 
		\and
		 S. Coelli\thanksref{INFNMI}
		 \and \\
		 O. Dorvaux\thanksref{Strasb}
		 \and 
		 S. Erturk\thanksref{Turk}
		 \and
		  G. De France\thanksref{GANIL} 
		  \and
		   C. Fransen\thanksref{IKPCol}
		  \and
		   A.  Goldkuhle\thanksref{IKPCol}
		  \and \\
		   J. Gr\k{e}bosz\thanksref{IFJPAN} 
		  \and
		   M.N. Harakeh\thanksref{KVI}
		   \and
			{\L}.W. Iskra\thanksref{IFJPAN,INFNMI}
			\and 
			B. Jacquot\thanksref{GANIL} 
			\and
			A. Karpov\thanksref{Dubna}
			\and \\
			M. Kici\`nska-Habior\thanksref{WarsUNI}
			\and
			Y. -H. Kim\thanksref{GANIL,ILL} 
			\and
			M. Kmiecik\thanksref{IFJPAN}
			\and
			A. Lemasson\thanksref{GANIL} 
			\and \\
			S.M. Lenzi\thanksref{UNIPD,INFNPD}
			\and
			M. Lewitowicz\thanksref{GANIL} 
			\and
			H. Li\thanksref{GANIL} 
			\and
			 I. Matea\thanksref{Orsay}  
			 \and
			 K. Mazurek\thanksref{IFJPAN}
			 \and \\
			C. Michelagnoli\thanksref{ILL} 
			\and
			M. Matejska-Minda\thanksref{WarsLAB,IFJPAN}
			\and 
			B. Million\thanksref{INFNMI}
			\and \\
			C. M\"{u}ller-Gatermann\thanksref{IKPCol,Argonne}
			\and
			V. Nanal\thanksref{Tata} 
			\and 
			P. Napiorkowski\thanksref{WarsLAB}
			\and
			D.R. Napoli\thanksref{LNL}
			\and
			R. Palit\thanksref{Tata}
			\and
			M. Rejmund\thanksref{GANIL} 
			\and
			Ch. Schmitt\thanksref{Strasb}
			\and
			M. Stanoiu\thanksref{IFIN-HH} 
			\and
			I. Stefan\thanksref{Orsay} 
			\and  \\
			E. Vardaci\thanksref{INFNNA} 
			\and
			B. Wasilewska\thanksref{IFJPAN}
			\and
			O. Wieland\thanksref{INFNMI} 
			\and \\
			M. Zi\k{e}bli\'{n}ski\thanksref{IFJPAN} 
			\and
			M. Zieli\'{n}ska\thanksref{IRFU} 
}

\thankstext{e1}{e-mail: Michal.Ciemala@ifj.edu.pl (corresponding author)}
\thankstext{e2}{e-mail: Silvia.Leoni@mi.infn.it (corresponding author)}
\thankstext{e3}{e-mail: Bogdan.Fornal@ifj.edu.pl (corresponding author)}

\authorrunning{Short form of author list} 

\institute{Institute of Nuclear Physics, PAN, 31-342 Kraków, Poland \label{IFJPAN}
		  \and
          Dipartimento di Fisica, Universit$\grave{a}$ degli Studi di Milano, I-20133 Milano, Italy \label{UNIMI}
           \and
           INFN Sezione di Milano, via Celoria 16, 20133, Milano, Italy \label{INFNMI}
           \and
          Institut f\"ur Kernphysik, Universit\"at zu K\"oln, 50937 Cologne, Germany \label{IKPCol}
           \and
          GANIL, CEA/DRF-CNRS/IN2P3, Bd. Henri Becquerel, BP 55027, F-14076 Caen, France \label{GANIL}
          \and
          CNRS/IN2P3, IPHC UMR 7178, F-67037 Strasbourg, France \label{Strasb}
           \and
          Nigde Omer Halisdemir University, Science and Art Faculty, Department of Physics, Nigde, Turkey \label{Turk} 
           \and
          University of Groningen, Groningen, the Netherlands \label{KVI}
          \and
          FLNR, JINR, 141980 Dubna, Russia \label{Dubna}
		   \and
		   Faculty of Physics, University of Warsaw, Warsaw, Poland \label{WarsUNI}
		   \and
		   Dipartimento di Fisica e Astronomia, Universit$\grave{a}$ degli Studi di Padova, I-35131 Padova, Italy \label{UNIPD}
		   \and
		   INFN Sezione di Padova, I-35131 Padova, Italy \label{INFNPD}
		   \and
          IPN Orsay Laboratory, Orsay, France \label{Orsay}
          \and
          Institut Laue-Langevin (ILL), Grenoble, France \label{ILL}
          \and
          Heavy Ion Laboratory, University of Warsaw, PL 02-093 Warsaw, Poland \label{WarsLAB}
          \and
         Physics Division, Argonne National Laboratory, Argonne, Illinois 60439, USA \label{Argonne}
         \and
          Tata Institute of Fundamental Research, Mumbai 400005, India \label{Tata}
          \and
          INFN Laboratori Nazionali di Legnaro, I-35020 Legnaro, Italy \label{LNL}
          \and
          IFIN-HH, Bucharest, Romania \label{IFIN-HH}
          \and
          Universit$\grave{a}$ degli Studi di Napoli and INFN sez. Napoli, Italy \label{INFNNA}
          \and
          RFU/CEA, Universit\'e Paris-Saclay, F-91191 Gif-sur-Yvette, France \label{IRFU}
}

\date{Received: date / Accepted: date}

\maketitle

\begin{abstract}
A novel Monte Carlo technique has been developed to determine lifetimes of excited states in the tens-to-hundreds femtoseconds range. The method is applied to low-energy heavy-ion binary reactions populating nuclei with complex velocity distributions. Its relevance is demonstrated in connection with the $^{18}$O(7.0 MeV/u) +  $^{181}$Ta experiment, performed at GANIL with the AGATA+VAMOS+PARIS setup, to study neutron-rich O, C, N, ... nuclei. Excited states in $^{17}$O and $^{19}$O, with known lifetimes, are used to validate the method over the $\sim$20-400 fs lifetime-sensitivity range. Emphasis is given to the unprecedented position resolution provided by $\gamma$-tracking arrays, which turns out to be essential for reaching the required accuracy in Doppler-shift correction, at the basis of the detailed analysis of $\gamma$-ray lineshape and resulting state lifetime determination. The technique is anticipated to be an important tool for lifetime investigations in exotic neutron-rich nuclei, produced with intense ISOL-type beams.
\end{abstract}
\section{Introduction}
\label{intro}
The study of exotic nuclei, {\it{i.e.}}, nuclear systems away from the valley of stability, is a central topic in modern nuclear physics. A detailed knowledge of their properties is needed to probe the evolution of the nuclear structure as a function of neutron and proton excess, and to understand the element abundances in the Universe. The heavy-element nucleosynthesis processes in stars \cite{Bur57}, as for example the r-process, are in fact to  large extent governed by the structural properties of the atomic nuclei involved \cite{Mol03}. Detailed experimental investigations are therefore needed in exotic regions of the nuclear chart, which are hard to reach by standard reaction mechanisms. 
In this context, low-energy binary collisions (which include multi-nucleon transfer and deep-inelastic reactions) \cite{Sch85,Kau61,Wil73,Zag14,Kar17,Ste18} are considered among the most favorable processes to populate yrast and near-yrast states in nuclei with large neutron excess, when high-intensity radioactive ISOL beams, presently under development, come into operation \cite{Kos12}. With the employment of powerful modern detection systems (for both particles and $\gamma$ rays), high-precision $\gamma$-spectroscopy measurements of very-exotic nuclei will become feasible, yielding nuclear structure information in terms of level energies, spins, parities, state lifetimes, etc.

In this paper, we present a novel approach to access nuclear state lifetimes in the tens-to-hundreds femtoseconds range. Our method makes use of low-energy binary heavy-ion collisions, where the complex structure of the product velocity distribution, caused by large energy dissipation  \cite{Sch85,Kau61,Wil73,Zag14,Kar17,Ste18}, does not allow to use standard Doppler-shift attenuation methods \cite{Nol79}. Such a short time range cannot be accessed by relativistic heavy-ion fragmentation either, for which the lower limit in lifetime determination is a few hundreds of femtoseconds \cite{Mor18}. 

The technique discussed in this work relies on the high-precision detection capabilities which are now reached with $\gamma$-ray tracking arrays, such as AGATA \cite{Akk12,Cle17,Kor20} and GRETINA \cite{GRETINA,Fal16}, coupled to powerful ancillary setups for heavy-ion identification, \textit{e.g.}, the VAMOS++ \cite{Rej11,Pul08}, PRISMA \cite{Ste02,Mon11}, FMA \cite{FMA} and S800 \cite{S800} spectrometers. The method has been first applied to extract the lifetime of the second 2$^+$ states in $^{16}$C and $^{20}$O, which have been predicted to be in the hundred-femtoseconds time range and to strongly depend on the three-body term of the nuclear interaction \cite{Cie20}. The experiment was performed at the Grand Acc\'el\'erateur National d'Ions Lourds (GANIL) in Caen, France, using the AGATA setup coupled to an early implementation of the PARIS scintillator array \cite{Maj09} and to the VAMOS++ heavy-ion spectrometer \cite{Rej11,Pul08}. In the measurement, light neutron-rich nuclei of  B, C, N, O and F were produced in low-energy binary processes induced by an $^{18}$O beam on a thick $^{181}$Ta target. In the following, the newly developed lifetime analysis will be discussed in details, in connection with this specific reaction case. 

The paper is organized as follows: the experimental setup is presented in Sec. \ref{sec:2}, while the analysis of the data is discussed in Sec. \ref{sec:3}, focusing on both the heavy-ion identification in the VAMOS++ magnetic spectrometer and the reconstruction of the associated Doppler-shift corrected $\gamma$-ray spectra, measured in the AGATA tracking array. Section \ref{sec:4} describes in details the new lifetime analysis method, based on a Monte Carlo simulation technique.  Selected excited states in $^{17}$O and $^{19}$O, with known lifetimes, are considered to validate the technique over the lifetime sensitivity range, {\it{i.e.}}, 20-400 fs. The impact of the high precision provided by the AGATA tracking array in identifying the $\gamma$-ray interaction point, which is essential for reaching the required accuracy in Doppler-shift corrections, is also discussed.

\begin{figure}[ht]
\centering
\resizebox{0.48\textwidth}{!}{\includegraphics{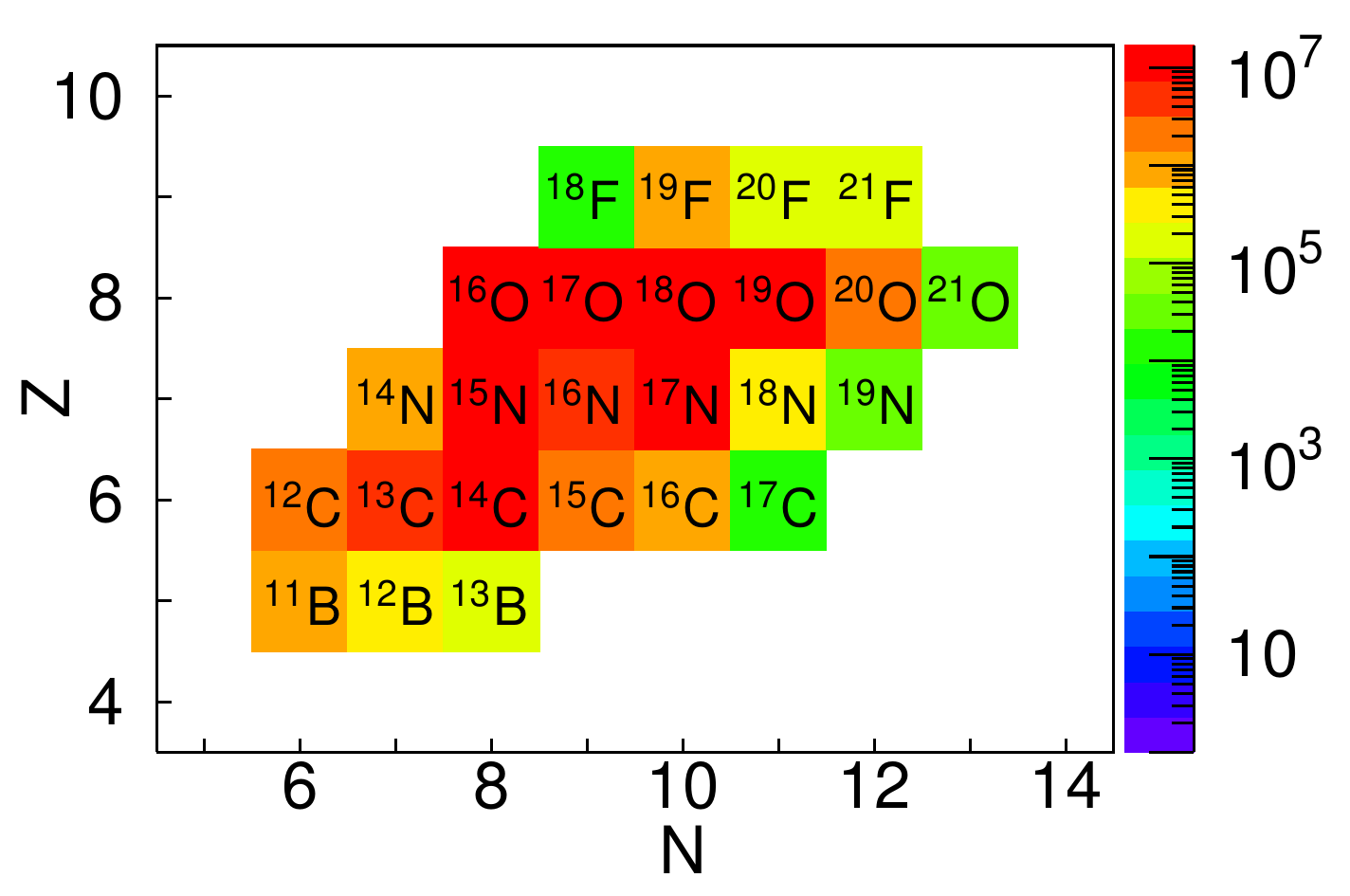}}
\caption{Population of detected and identified ions in the $^{18}$O (126 MeV) + $^{ 181}$Ta collision, investigated at GANIL with the AGATA+PARIS+VAMOS setup \cite{Cie20}. } 
\label{population}
\end{figure}

\section{Experiment and Setup}
\label{sec:2}

In the GANIL experiment, a beam of $^{18}$O at 126 MeV ({\it{i.e.}}, 7.0 MeV/u) impinging on a thick $^{181}$Ta target (6.64 mg/cm$^2$) was employed to induce direct transfer and deep-inelastic reactions producing a variety of neutron-rich nuclei, from B (Z=5) to F (Z=9), as shown in Fig. \ref{population} \cite{Cie20}. The beam energy at the center of the target was $\sim$116 MeV, {\it{i.e.}}, $\sim$50$\%$ above the Coulomb barrier, and projectile-like products had velocities of v/c $\sim$10$\%$.

\begin{figure}[ht]
\centering
\resizebox{0.48\textwidth}{!}{\includegraphics{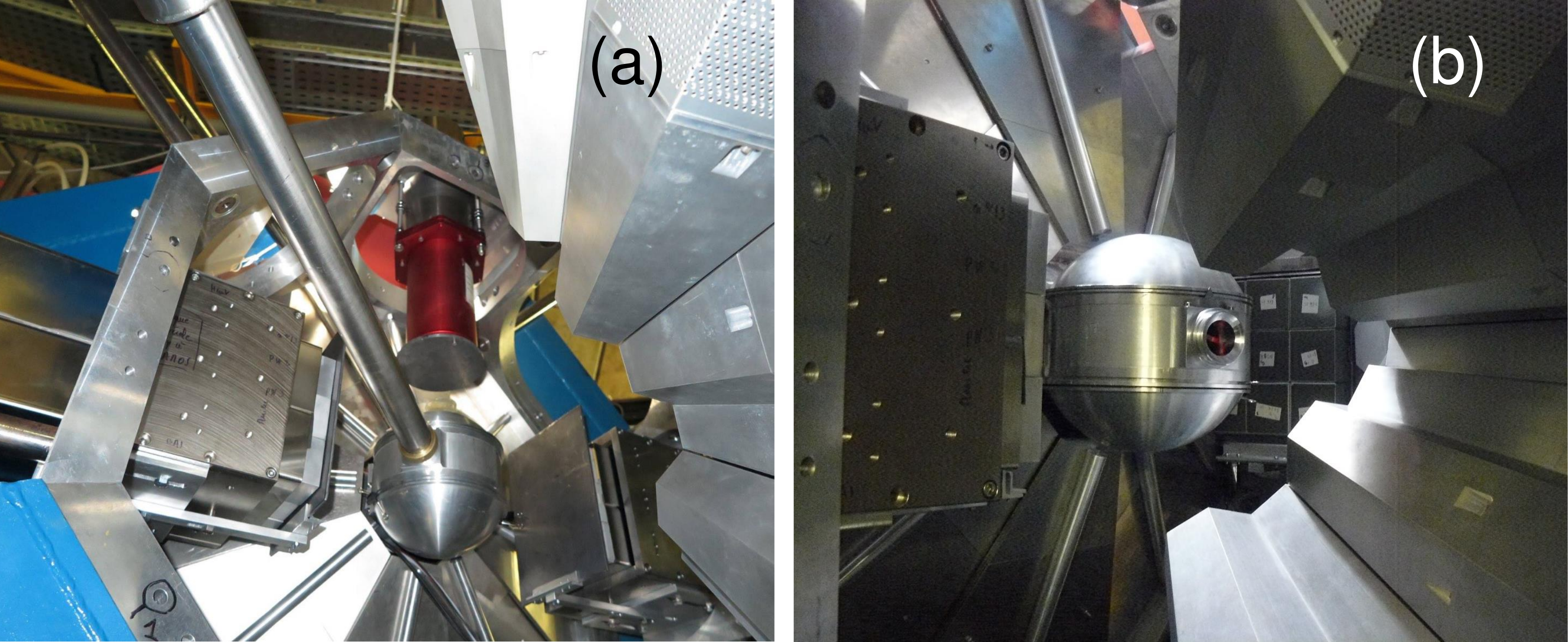}}
\caption{Pictures of the experimental setup. Left panel: overall view of the AGATA and scintillator arrays. The spherical scattering chamber is visible in the middle, surrounded by the two PARIS clusters, the two large-volume LaBr$_3$:Ce scintillators (one at the top, in red) and the AGATA detectors (on the right). Right panel: closer view of the AGATA detectors (on the right), the two PARIS clusters (without front absorbers), and the scattering chamber in the middle. } 
\label{setup}
\end{figure}

Following the reaction, the $\gamma$ rays emitted by the excited nuclei were detected by the AGATA tracking array \cite{Akk12,Cle17,Kor20}, consisting of 31 segmented High Purity Germanium (HPGe) detectors, coupled to the PARIS scintillators array \cite{Maj09}, with two complete clusters of nine phoswich detectors each, plus two large-volume (3.5 "$\times$ 8 ") LaBr$_3$:Ce scintillators \cite{Gia13,Gos18}. Pictures of the experimental setup are given in Fig. \ref{setup}.

The projectile-like products were detected in the VAMOS++ magnetic spectrometer \cite{Rej11,Pul08}, placed at the reaction grazing angle of 45$^{\circ}$, relative to the beam direction, with an aperture of $\theta$ = $\pm$ 6$^{\circ}$  and aligned with the center of AGATA. The PARIS array was placed at 90$^{\circ}$  with respect to the VAMOS++ axis, while AGATA covered the angular range between   $\sim$115$^{\circ}$  and  $\sim$175$^{\circ}$.

\section{Data Processing}
\label{sec:3}

In the following Sections, the processing of the data, collected in the $^{18}$O+$^{181}$Ta experiment, is discussed in detail. Section \ref{sec:3.1} is devoted to the VAMOS++ magnetic spectrometer, which allows to identify the atomic number Z and the mass A of the ions, and to precisely reconstruct their trajectories. Fine corrections, which can be applied to the masses and beam-spot reconstruction, taking advantage of recently implemented VAMOS++ entrance detectors and of the fast scintillators of PARIS, are discussed in Sec. \ref{subsubsec:3.1.1}. Section \ref{subsec:3.2} focuses instead on the offline processing of the AGATA data. 

\subsection{Ions identification and trajectory reconstruction in VAMOS++}
\label{sec:3.1}

\begin{figure}[ht]
\centering
\resizebox{0.48\textwidth}{!}{\includegraphics{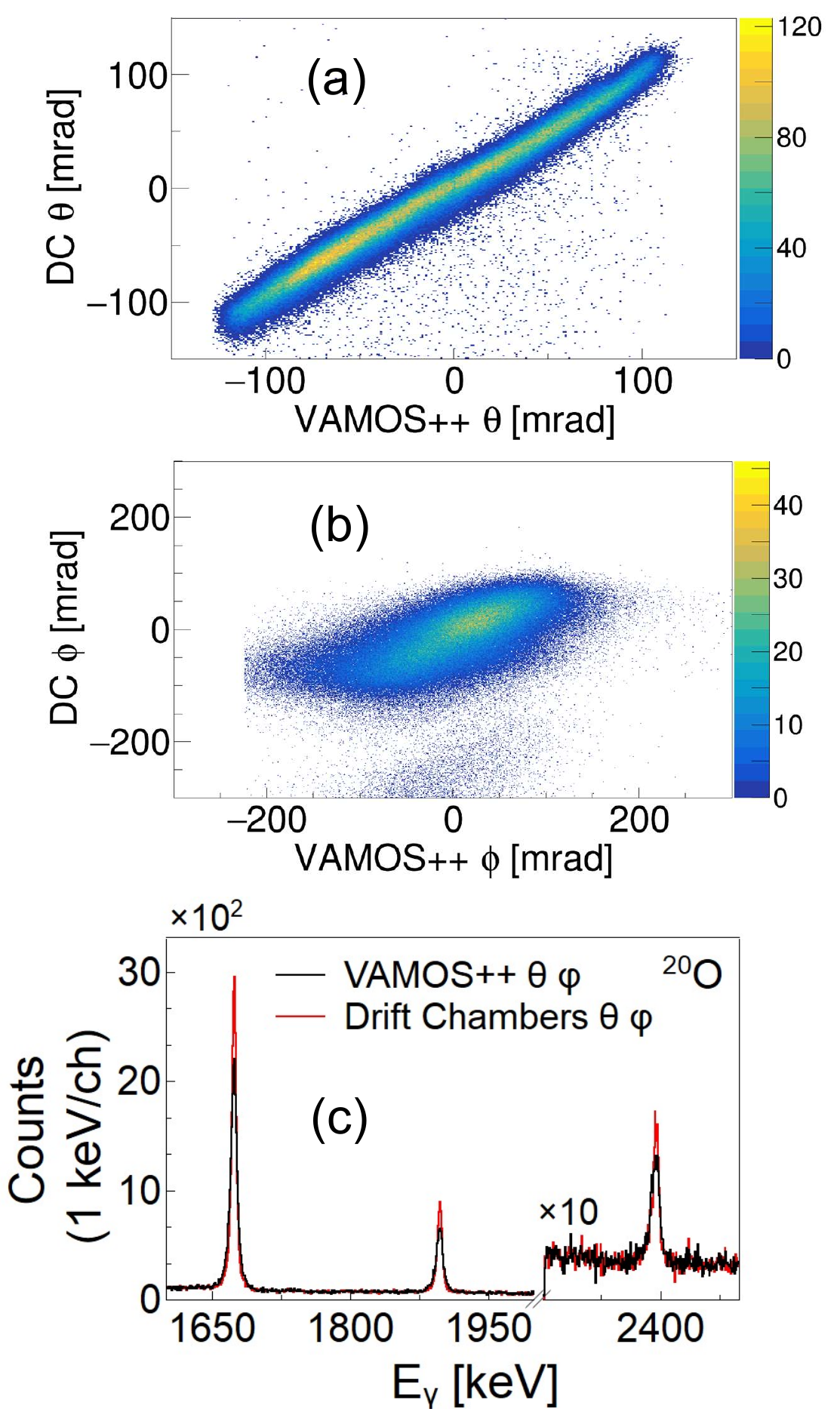}}
\caption{Panel (a) and (b): drift chambers (DCs)  vs. VAMOS++ reconstructed $\theta$ and $\phi$ angles of the recoiling ions, respectively. Panel (c): comparison between Doppler-shift corrected $^{20}$O $\gamma$-ray spectra measured in AGATA, based on the reconstructed $\theta$ and $\phi$ angles from the VAMOS++ focal-plane detectors only (black), or on the use of the entrance drift chambers (red). The FWHM of the peaks improves by a factor of 1.4 for $\gamma$-ray energies around 1.6-1.9 MeV.} 
\label{VAMOSangle}
\end{figure} 

The focal plane of the VAMOS++ spectrometer consisted of: i) four drift chambers, providing the x$_f$, y$_f$ position information for the reconstruction of the $\theta_f$, $\phi_f$ angles of the reaction product, ii) a segmented ionization chamber, divided into six columns and four rows, for measuring the ion energy loss $\Delta$E, and iii) one long plastic scintillator at the end of the focal plane, giving the trigger signal, the particle residual energy E and the time with respect to the cyclotron radiofrequency (RF).
Two additional pairs of drift chambers (DC) were also placed at the entrance of VAMOS++, at 20 cm distance from the target, in order to accurately determine the $\theta$ and $\phi$ angles of emission of the light ions, with good efficiency \cite{DCRep}. They also significantly  improved the $\gamma$-ray Doppler-shift correction. Figures \ref{VAMOSangle}(a) and \ref{VAMOSangle}(b) show comparisons of the reconstructed $\theta$ and $\phi$ angles, respectively, using VAMOS++ and the entrance DCs. It is seen that the $\theta$ angle is well determined by VAMOS++ (a), while this is not the case for the  $\phi$  angle (b). Panel (c) displays a $\gamma$ spectrum of $^{20}$O, as measured in AGATA, obtained after applying a Doppler-shift correction in which the ions direction reconstruction was based on the VAMOS++ focal-plane detectors only (black line), or on the use of the entrance drift chambers (red line).  We remark that a similar improvement of the Doppler reconstruction was also discussed in Ref. \cite{Van16}, where  a Multi-wire chamber was used for heavy and slow ions detection. In the following, the more accurate angles from the entrance DCs will be used for the ion direction entering the Doppler-shift correction, rather than the focal-plane reconstructed ones (see Sec. \ref{subsubsec:3.2.1}).

\begin{figure}[ht]
\centering
\resizebox{0.48\textwidth}{!}{\includegraphics{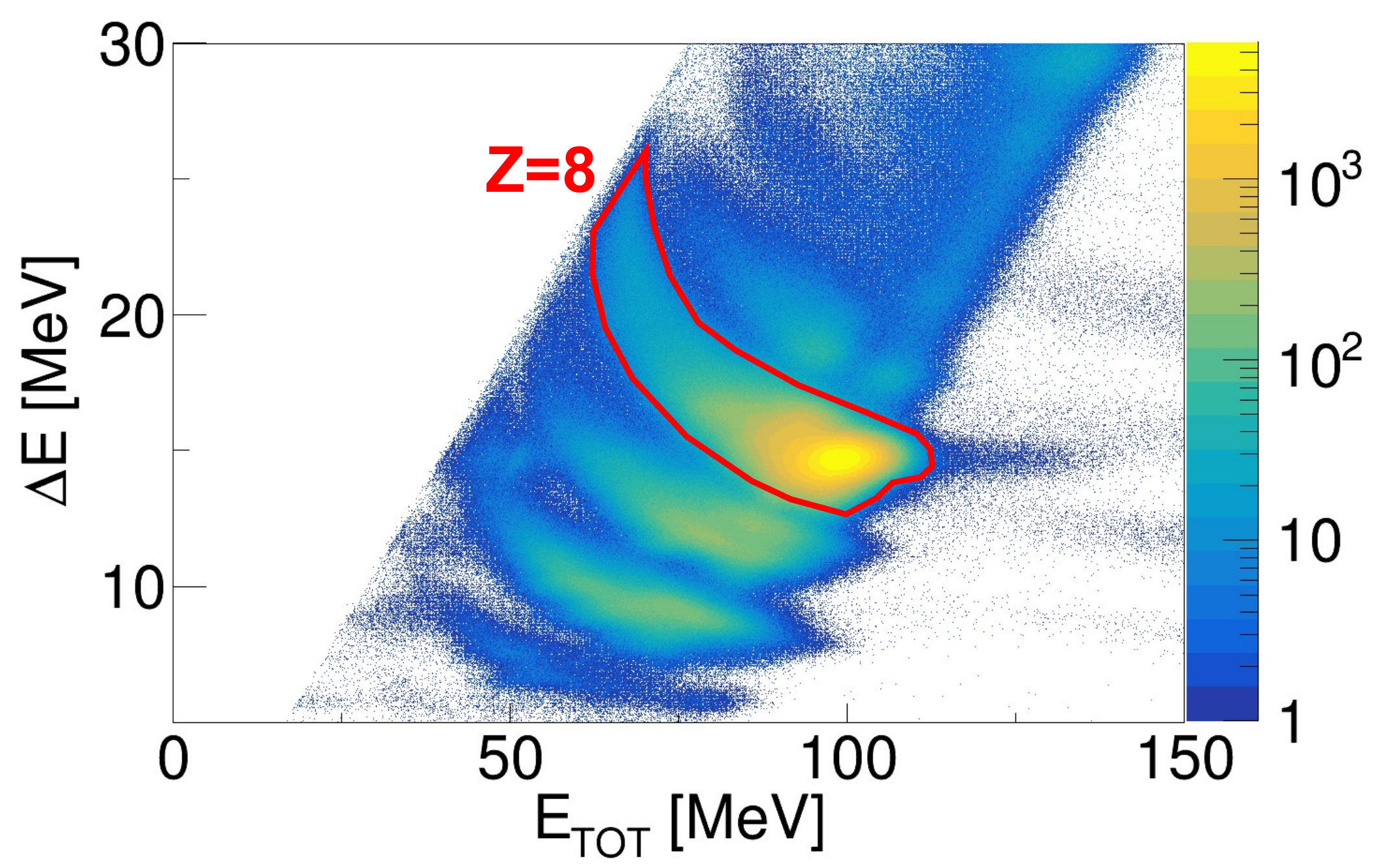}}
\caption{$\Delta$E vs. E$_{TOT}$ two-dimensional histogram, where $\Delta$E and E$_{TOT}$ are the energy loss and total ion energies measured by the ionization chambers and focal-plane plastic detectors of the VAMOS++ magnetic spectrometer, in the $^{18}$O (126 MeV) + $^{ 181}$Ta experiment. As an example, the region corresponding to Z=8 is encircled by a red line \cite{Cie20}. } 
\label{VAMOS_Z}
\end{figure}

\begin{figure}[ht]
\centering
\resizebox{0.48\textwidth}{!}{\includegraphics{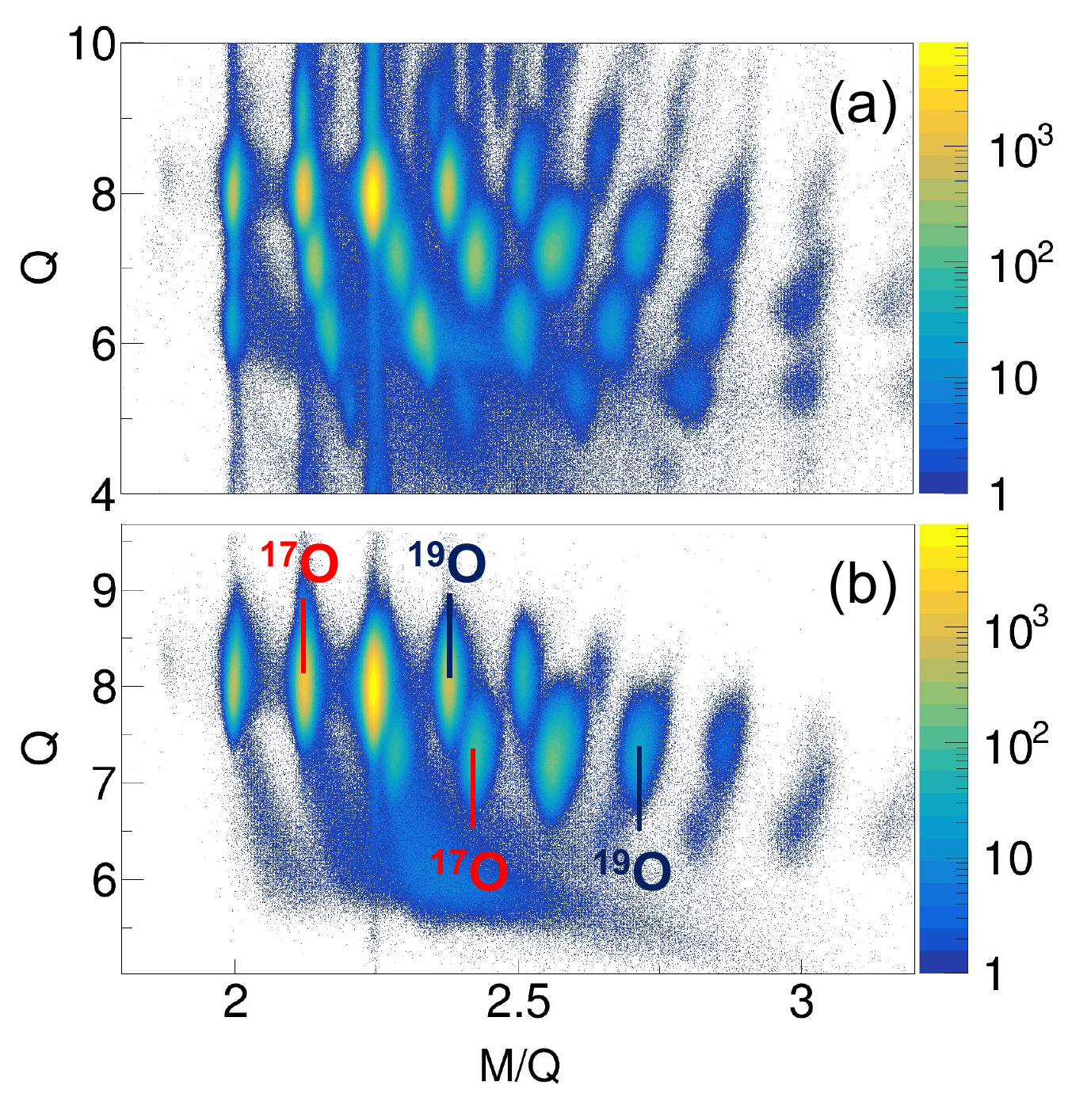}}
\caption{Plot of the ion charge Q vs. M/Q for all ions (a) and for the selection of oxygen (Z=8) (b), with labels pointing to $^{17}$O and $^{19}$O. Data refer to the  $^{18}$O (126 MeV) +  $^{181}$Ta experiment  \cite{Cie20}. } 
\label{VAMOS_Mass}
\end{figure}

The ion velocity $v$ was then obtained from the relation $v = D/T$, where $D$ and $T$ are, respectively, the ion path length and the time-of-flight (ToF), from the target to the focal-plane plastic detector.  The ToF was calculated using the focal-plane plastic detector time signal and the RF signal of the cyclotron (with a period of 102 ns), as a reference.  The ion mass $M$ was reconstructed by employing the standard VAMOS++ analysis procedure \cite{Rej11,Pul08,LibVAMOS}, and the ion atomic number Z was determined from the correlation between the energy loss $\Delta$E and the total energy E$_{TOT}$, as shown in Fig. \ref{VAMOS_Z}. Figure \ref{VAMOS_Mass} shows the plot of the product-charge state $Q$ versus $M$ for all $Z$ (panel (a)),  and for $Z$ = 8 (panel (b)).

\subsubsection{Improvement on product mass and trajectory reconstruction}
\label{subsubsec:3.1.1}
\begin{figure}[ht]
\centering
\resizebox{0.4\textwidth}{!}{\includegraphics{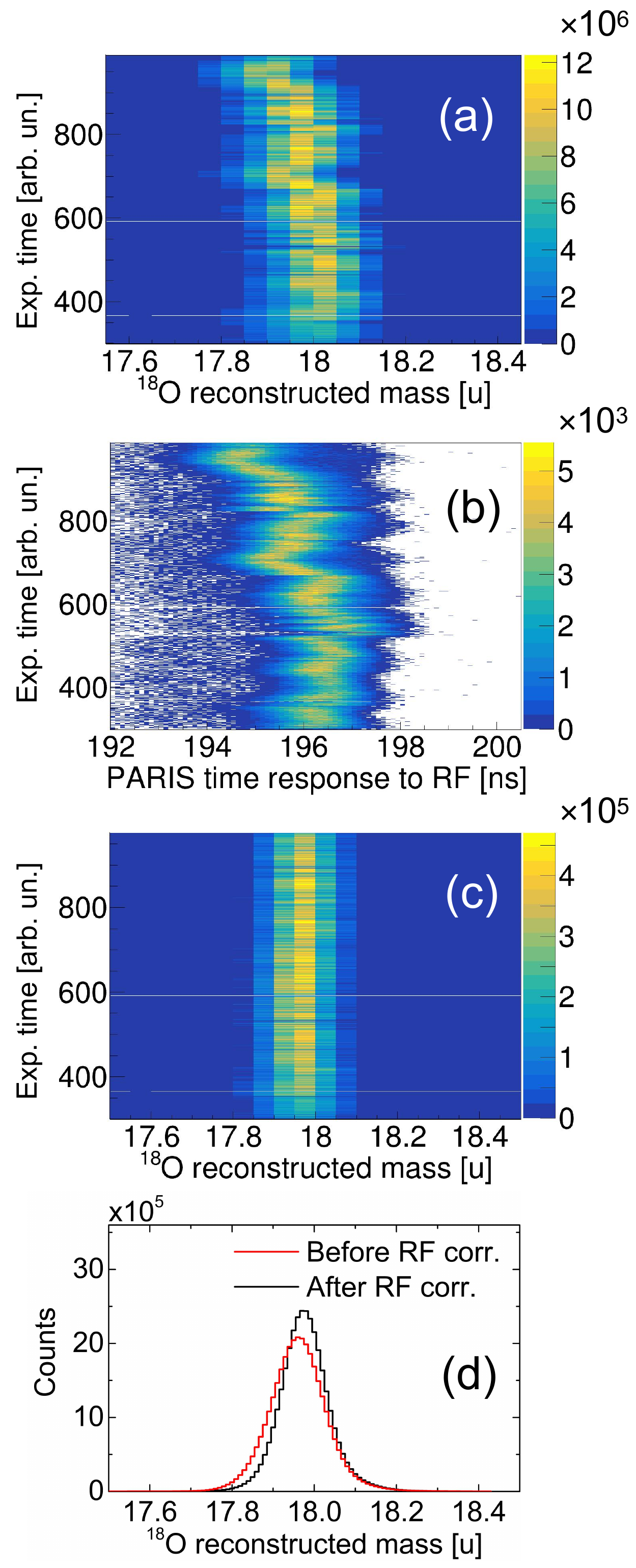}}
\caption{: (a) Reconstructed $^{18}$O mass, as a function of the experiment duration time. (b) PARIS time signal with respect to the cyclotron RF. (c) Reconstructed $^{18}$O mass after the RF drift correction, estimated from the PARIS time drift with respect to the RF (panel (b)). (d) $^{18}$O reconstructed mass before (red) and after (black) correction for the RF drift.  The FWHM improves by a factor of 1.2. } 
\label{VAMOS_Mass_corr}
\end{figure}

Two corrections were introduced to the aforementioned "standard" VAMOS++ identification method, in order to improve the determination of ion mass and trajectory. First, a check was done on the stability of the reconstructed masses M$_r$ in the course of the $^{18}$O (126 MeV) +  $^{181}$Ta measurement \cite{Cie20}. Panel (a) of Fig. \ref{VAMOS_Mass_corr} shows the evolution of the reconstructed $^{18}$O ion mass, as a function of the experiment duration time. A clear drift is visible, in phase with the drift observed for the time signal of the PARIS scintillators, with respect to the cyclotron radiofrequency (RF) (panel (b)). It follows that a correction to the drift in the ion time-of-flight T (on which the mass reconstruction is based) can be extracted from the PARIS time vs. RF drift. As shown in Figure \ref{VAMOS_Mass_corr}(c), this leads to a significantly improved stability of the reconstructed masses and to an overall improved mass resolution (panel (d)).

\begin{figure}[ht]
\centering
\resizebox{0.48\textwidth}{!}{\includegraphics{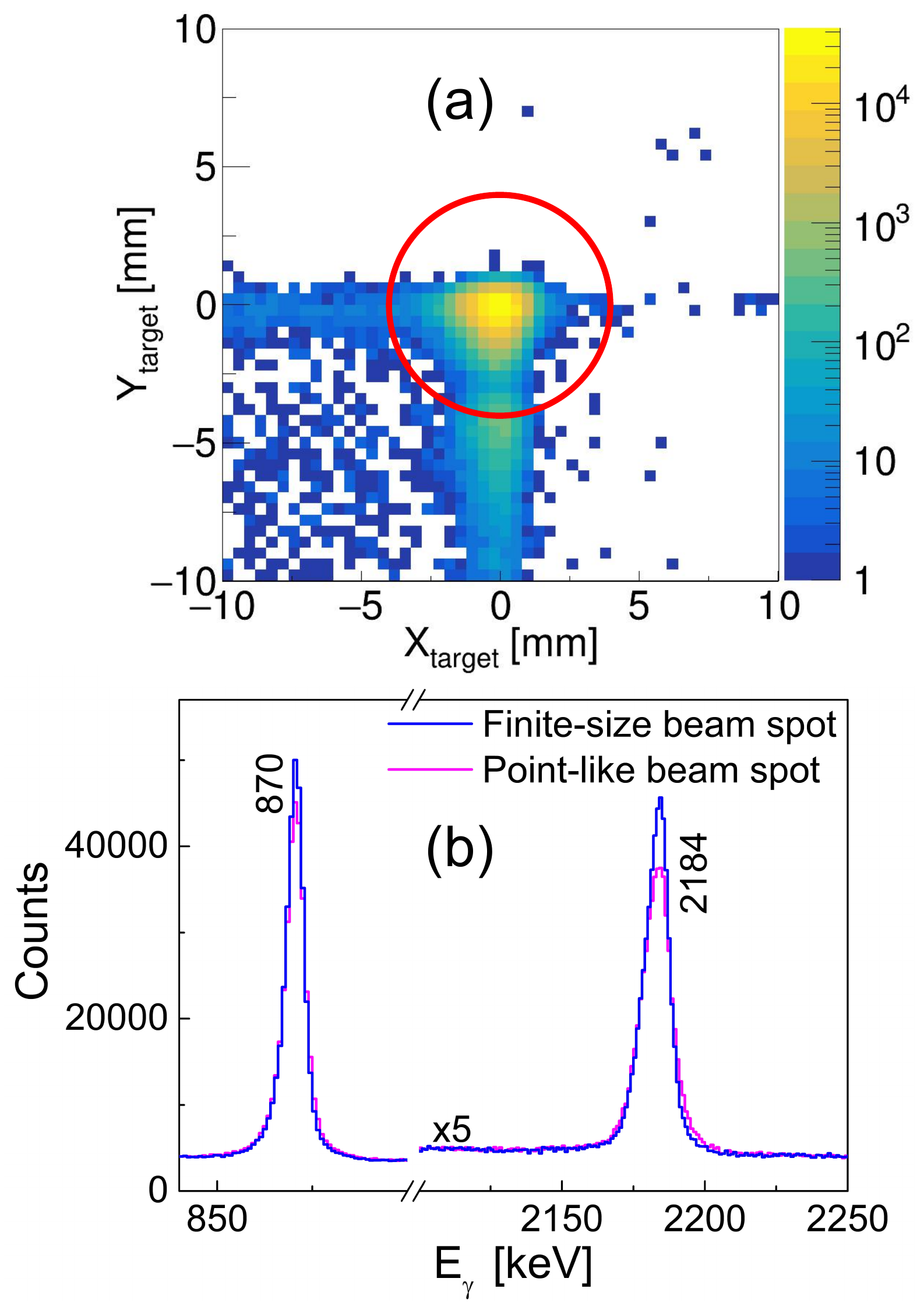}}
\caption{Panel (a): Beam-spot reconstruction obtained by using information on the x and y positions from the entrance drift chamber detectors of VAMOS++. The distribution has a Gaussian-like shape with $\sigma_x$ = 0.5 mm and $\sigma_y$  = 0.4 mm. The red circle (with a 4 mm radius) delimits the acceptance area for the event reconstruction. (b): AGATA Doppler-shift corrected $\gamma$-ray energy spectrum of  $^{17}$O, as obtained assuming a point-like (pink) or a finite-size reconstructed (blue) beam spot.} 
\label{Beam_spot}
\end{figure}

Second, for what concerns the ion trajectory reconstruction, it is found that a significant improvement in the $\gamma$-ray Doppler-shift correction is obtained by considering a finite-size instead of a point-like beam spot. The position of the beam on target can in fact be deduced by considering the ion track direction, as measured  by the two pairs of Drift Chambers at the entrance of VAMOS++. 
Considering the geometry of the setup ({\it{i.e.}}, target-detector distances) and this ion-reconstructed direction, a Gaussian-like beam-on-target distribution was obtained, with $\sigma_x$=0.5 mm and $\sigma_y$=0.4 mm, as shown in Fig. \ref{Beam_spot}(a). In the data analysis, only ion trajectories originating within a 4 mm distance from the beam-spot center were considered, leading to rejection of wrongly reconstructed trajectories and improving the $\gamma$-ray Doppler-shift correction, in comparison with a point-like beam-spot assumption (see Fig. \ref{Beam_spot}(b)). 

\begin{figure}[ht]
\centering
\resizebox{0.48\textwidth}{!}{\includegraphics{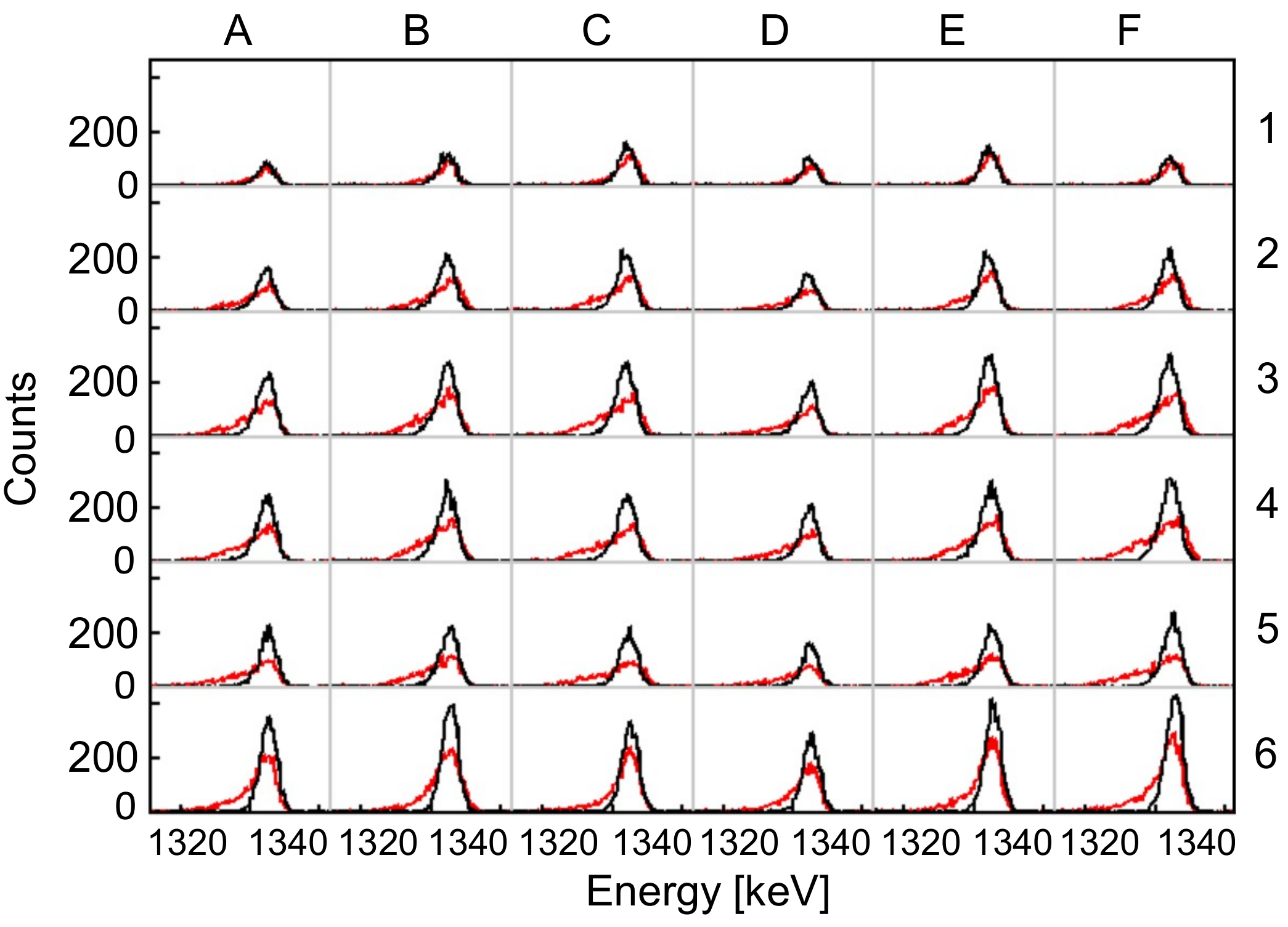}}
\caption{Examples of energy lineshapes for the 1332-keV $^{60}$Co $\gamma$-ray from the 36 segments of  crystal 10B of AGATA, before (red) and after (black) the neutron-damage correction \cite{Bru13}. The source data refers to a calibration run performed for the $^{18}$O+$^{181}$Ta experiment. See also Fig. \ref{AGATA_sensitivity} for segments label code.} 
\label{Neutron_damage}
\end{figure}

\subsection{Offline processing of the AGATA data}
\label{subsec:3.2}
The most accurate treatment of the information from the AGATA $\gamma$-tracking array is crucial for a precise $\gamma$-ray lineshape analysis (see Sec. \ref{sec:4.5}). A data replay was therefore performed offline, using the NARVAL data-acquisition system emulator \cite{Gra05,NARVAL}: all the files containing the electronic traces from each AGATA crystal were processed, the Pulse Shape Analysis (PSA) and the matching of the data from different crystals was repeated, as well as the merging of the events from AGATA and ancillary detectors. The energy and direction of the interacting $\gamma$ ray in AGATA was reconstructed by the combined use of the PSA \cite{Ven04,Lew19} and of the Orsay Forward Tracking (OFT) algorithm \cite{Lop04}, which  allow to reach a position resolution of the order of 4 mm FWHM. During the offline data replay, crosstalk and neutron-damage corrections were applied, following the procedures described in Refs. \cite{Bru09,Bru13}. Figure \ref{Neutron_damage} illustrates the impact of the neutron-damage corrections on the 36 segments energy spectra of crystal 10B of AGATA, which results in a clear improvement in the peak symmetry and corresponding energy resolution. 

\begin{figure}[ht]
\centering
\resizebox{0.43\textwidth}{!}{\includegraphics{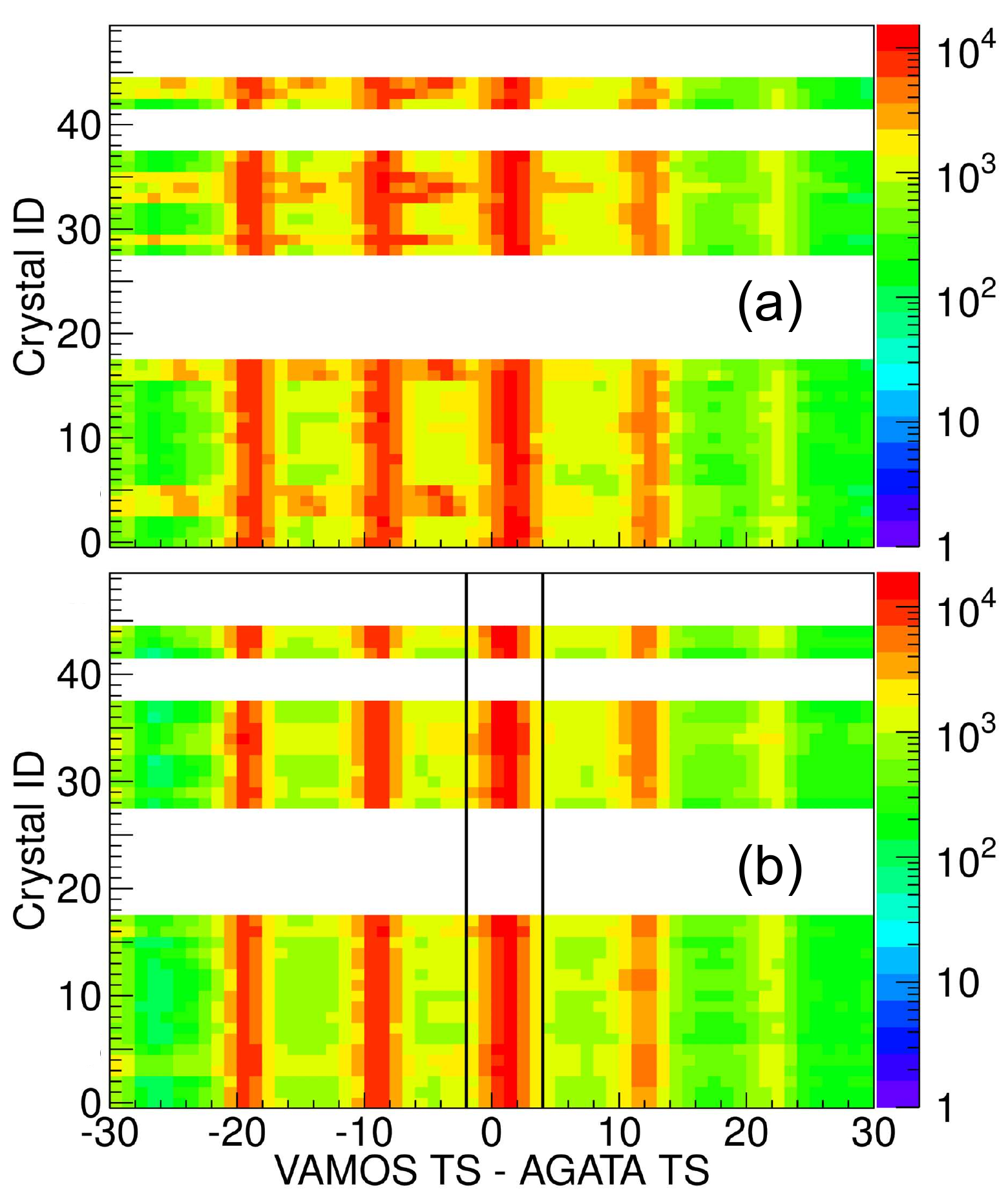}}
\caption{Examples of two-dimentional histograms showing the AGATA crystal identification number (Crystal ID, y axis) vs. timestamp difference between VAMOS++ and AGATA (TS, x axis), before (a) and after (b) the time-alignment procedure. White gaps correspond to missing detectors, black lines in (b) indicate the prompt coincidence gate used in the final analysis.  } 
\label{Time_stamp}
\end{figure}

The offline data replay was also needed for statistics recovery from missing or broken crystal segments, timestamp alignment and energy-calibration improvements. Figure \ref{Time_stamp} shows an example of the timestamp alignment of the AGATA crystals. By setting stringent gates on the timestamp difference between AGATA and VAMOS++  (black lines in panel (b)), an improvement of the peak-to-background ratio of a factor $>$ 2 was achieved, with a loss of counts in the photopeak less than 4$\%$

\begin{figure}[ht]
\centering
\resizebox{0.45\textwidth}{!}{\includegraphics{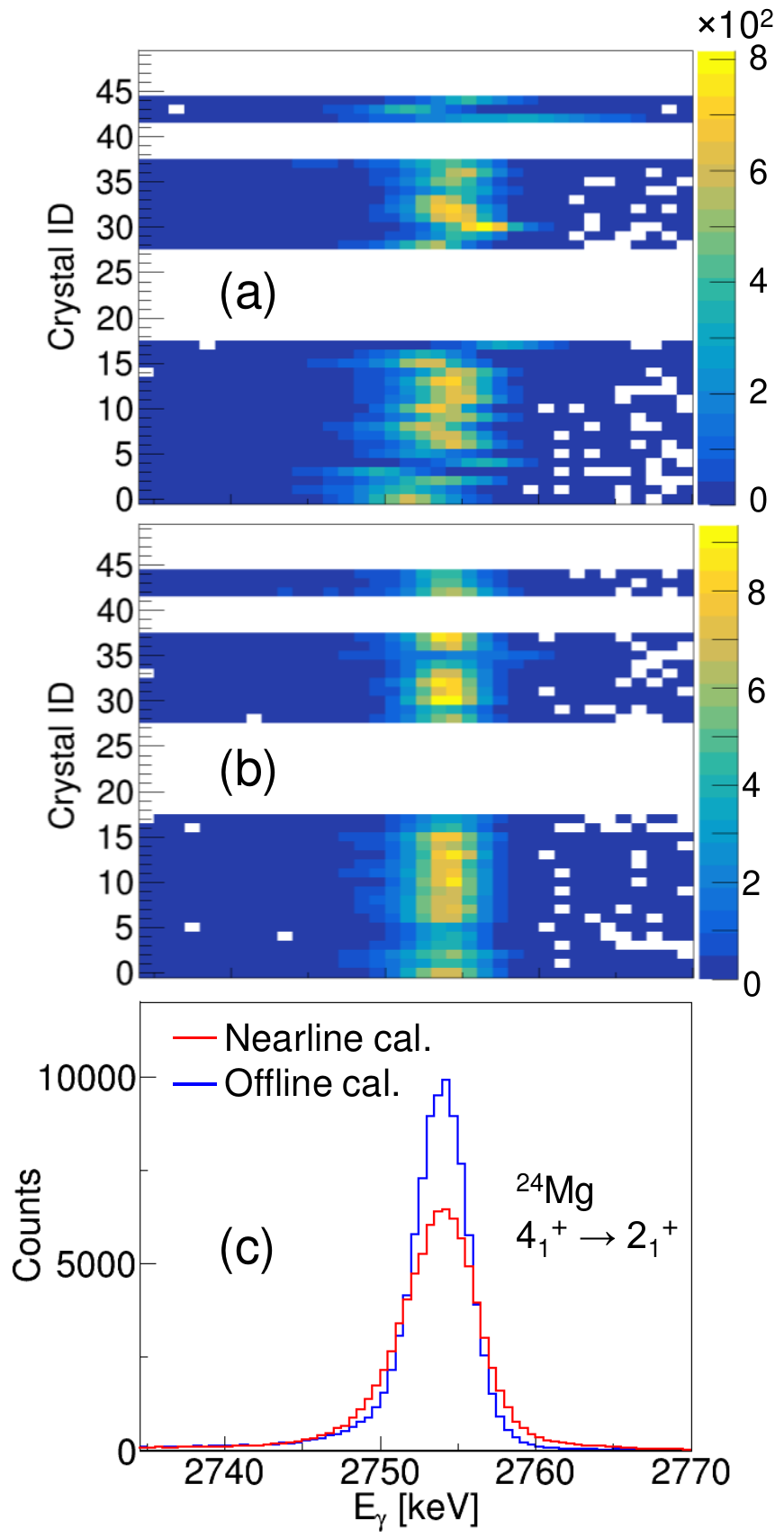}}
\caption{Examples of two-dimensional histograms showing the AGATA crystal identification number (Crystal ID, y axis) vs. energy spectra from calibration sources, in the region around the 2754-keV line from $^{24}$Mg, before (a) and after (b) the energy-calibration correction procedure.  Panel (c) shows the corresponding energy spectra, summed over all AGATA crystals, from which an improvement of a factor of 1.4 is deduced in the FWHM (see text for details). } 
\label{Calibration}
\end{figure}

An accurate energy calibration was then applied to the core signals of the AGATA detectors, using $\gamma$-ray lines from $^{152}$Eu, below 1.5 MeV, and radiation coming from $^{208}$Pb and $^{24}$Mg, present in the surrounding materials, in the 2-3 MeV energy region.  After the calibration process, discrepancies between tabulated and calibrated energies  below 0.2 keV were obtained for most of the detectors, with only 4 detectors having discrepancies around 0.5 keV. Figure \ref{Calibration} shows examples of two-dimentional histograms of AGATA energy spectra from calibration sources, before (a) and after (b) the energy-calibration procedure, in the region of the 2754-keV line from $^{24}$Mg. As shown in panel (c), an overall improvement of a factor of 1.4 is obtained in the FWHM of the 2754-keV line.

The energy calibration  was improved by forcing the summed energy measured in the crystal segments to be equal to the one measured in the core, for each $\gamma$-ray (the ForceSegmentsToCore option of the AGATA software package was used \cite{AGATA}). For one crystal (ID. 42), the core signal was degraded, therefore the individual segments were used. After this procedure, some missing energy could be recovered, resulting in a further reduction of the left-side tail of the energy peaks. 

Possible energy gain drifts with time were also checked using calibration sources and no appreciable drift was observed.

\subsubsection{Doppler-shift Correction }
\label{subsubsec:3.2.1}

\begin{figure}[ht]
\centering
\resizebox{0.38\textwidth}{!}{\includegraphics{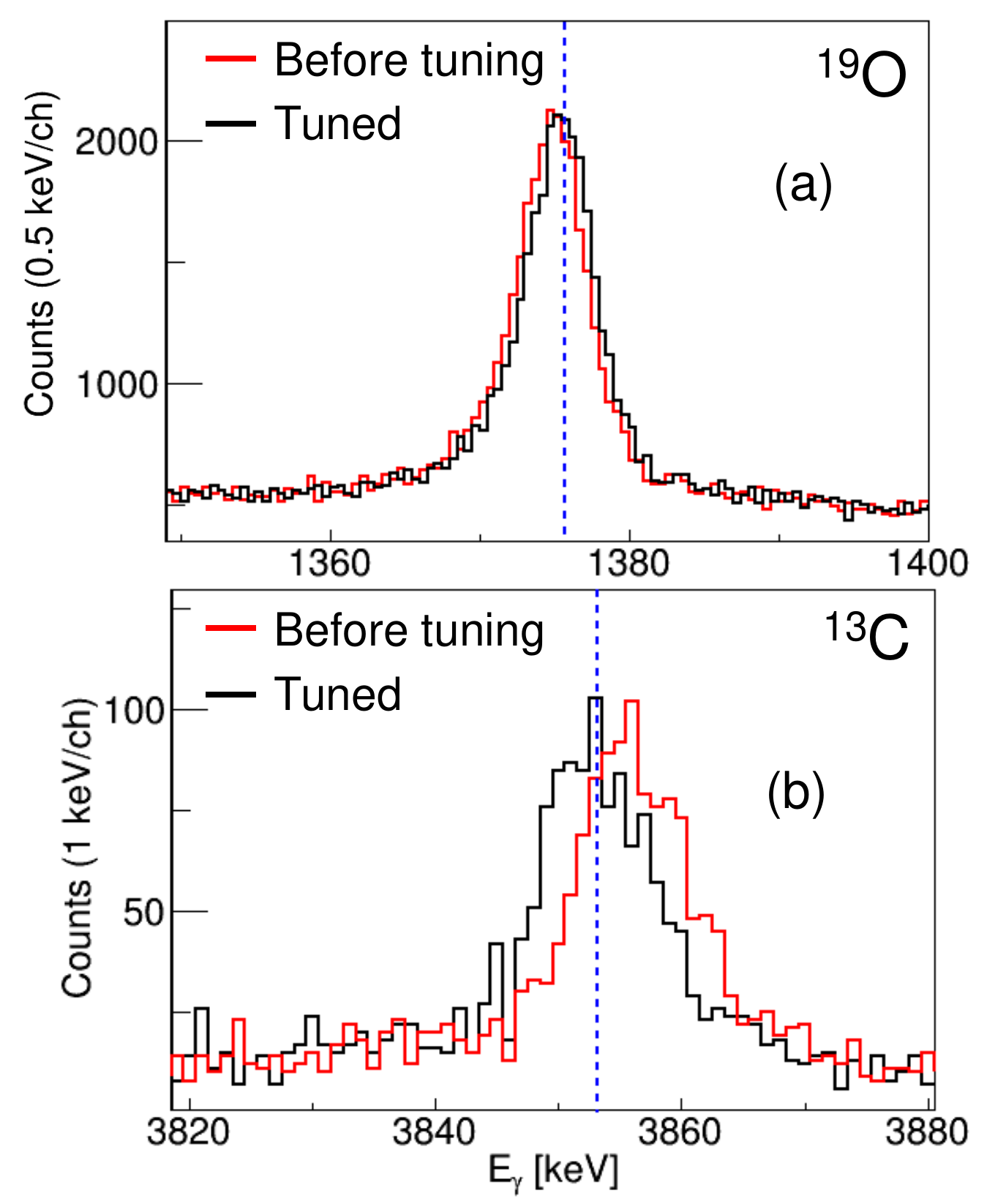}}
\caption{Effect of the fine-tuning of the ion velocity $v$, measured in VAMOS++ entrance DC's detectors, on the Doppler-shift corrected 1375.80(8) keV and 3853.170(22) $\gamma$-rays of $^{19}$O (a) and $^{13}$C (b), respectively. The corresponding changes in velocity are 0.4$\%$ and 0.7$\%$. Dashed blue lines indicate the nominal $\gamma$-ray energies \cite{Fir16,NNDC}.} 
\label{fine_tuning}
\end{figure}

After merging the AGATA and ancillaries events, via timestamps correlations, the Doppler-shift correction was applied to the $\gamma$-ray energies according to the formula:  
\begin{center}
\begin{equation}
E_{\gamma}=  \frac{E_{\gamma_0}}{\gamma(1-\beta cos(\theta))}
\label{Doppler}
\end{equation}
\end{center}

where  $E_{\gamma_0}$  is the measured $\gamma$-ray energy in the laboratory frame. $\theta$ is the angle between the direction of the reaction product (measured in the entrance Drift Chambers) and the emitted $\gamma$-ray direction, which is calculated by considering the $\gamma$ emission from the target center and the first interaction point in AGATA (see discussion in Sec. \ref{sec:4.5}). No appreciable difference is observed in the Doppler-shift correction, if the $\gamma$ direction is calculated considering the reconstructed emission point in the target (see Fig.  \ref{Beam_spot}).  In Eq. \ref{Doppler}, the $\beta$ and $\gamma$ relativistic terms are calculated by using a reconstructed ion velocity which also takes into account the energy loss in the entrance drift chambers of  VAMOS++ (see Sec. \ref{sec:3.1}). This energy loss is determined, at first, via the LISE code \cite{Tar08}, assuming the stopping power for oxygen ions, {\it{i.e.}}, beam-like reaction products. The most precise Doppler-shift correction is then achieved by a fine tuning of the velocity $v$, which is based on the reconstruction of Doppler-shift corrected $\gamma$-ray energies for known transitions de-exciting long-lived excited states (larger than 1 ps). Figure \ref{fine_tuning} shows the effect of velocity correction on the 1375.80(8) keV $\gamma$-ray in $^{19}$O (a) and on the 3853.170(22) keV $\gamma$-ray in $^{13}$C (b) (with nominal values \cite{Fir16,NNDC} indicated by dashed-blue lines). The corresponding changes in velocity are 0.4$\%$ for oxygen and 0.7$\%$ for carbon ions. 

\begin{figure}[ht]
\centering
\resizebox{0.40\textwidth}{!}{\includegraphics{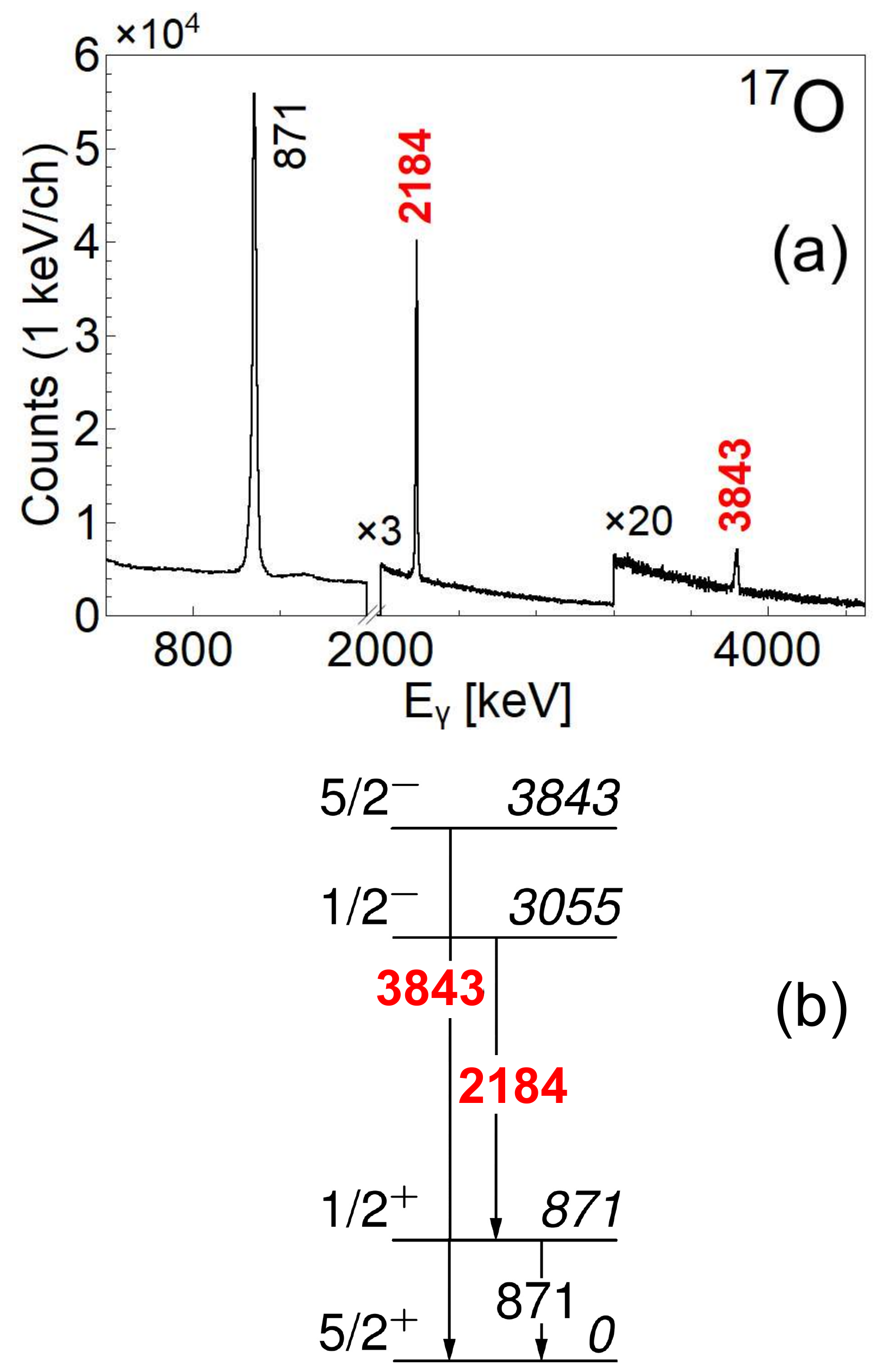}}
\caption{$^{17}$O  Doppler-shift corrected AGATA $\gamma$-ray spectrum (a) and corresponding level scheme (b) \cite{Cie20}. Transitions marked in red bold are considered in the lifetime analysis discussed in this work.} 
\label{17O}
\end{figure}

\begin{figure}[ht]
\centering
\resizebox{0.40\textwidth}{!}{\includegraphics{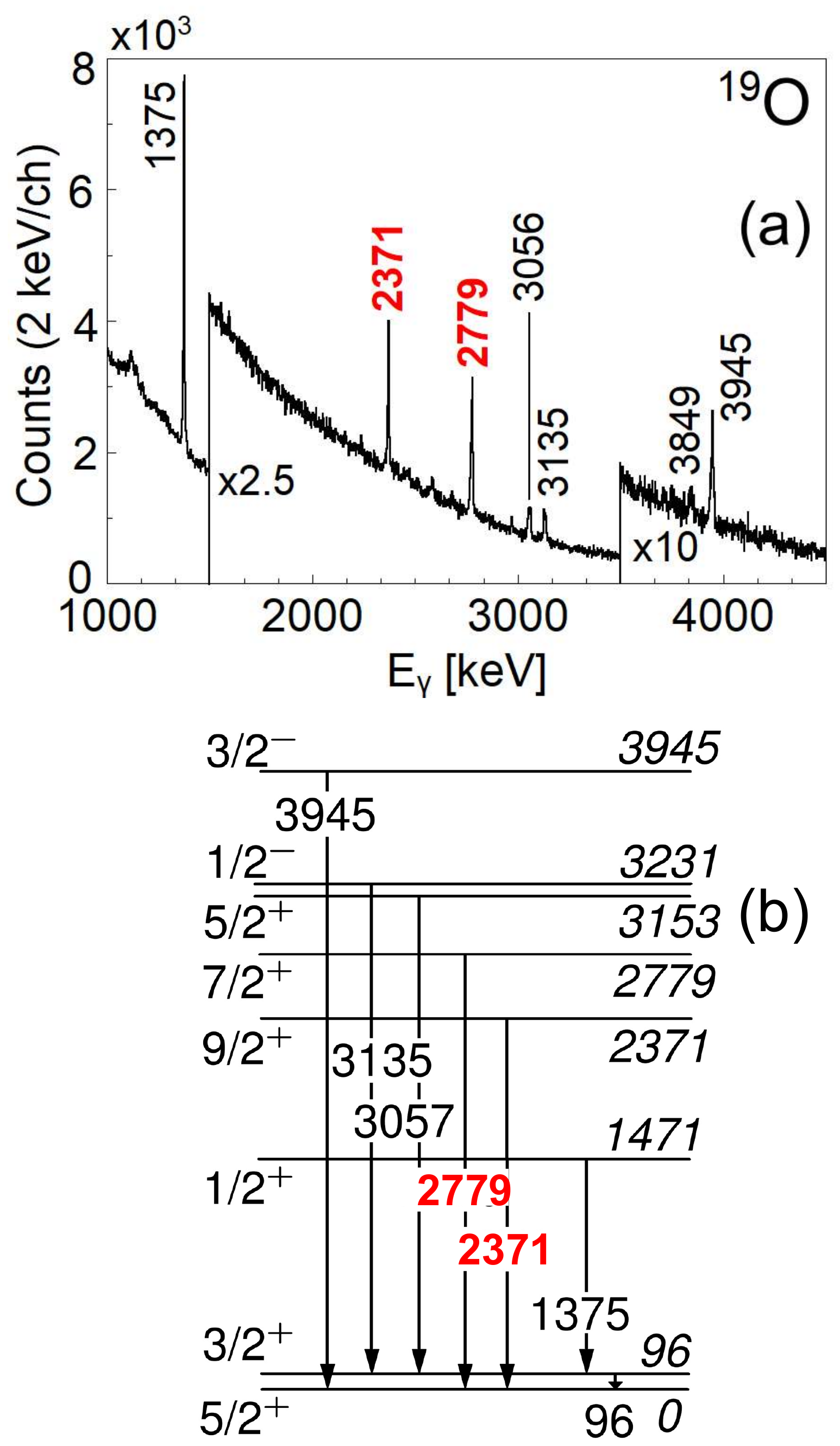}}
\caption{$^{19}$O  Doppler-shift corrected AGATA $\gamma$-ray spectrum (a) and corresponding level scheme (b) \cite{Cie20}. Transitions marked in red bold are considered in the lifetime analysis discussed in this work. } 
\label{19O}
\end{figure}

Following the procedure described in the previous sections and after applying the Doppler-shift correction here described, ion-gated  Doppler-shift corrected $\gamma$-ray spectra were obtained. As examples, Figures \ref{17O} and \ref{19O} show portions of Doppler-shift corrected $\gamma$-ray spectra of $^{17}$O and $^{19}$O, as measured by the AGATA array, and the corresponding level schemes.  Transitions of 2184- and 3843-keV energy in $^{17}$O, and 2371 and 2779 keV in $^{19}$O will be later considered for lifetime analyses.

\section{Monte Carlo Simulations for lifetime analysis}
\label{sec:4}

The present section describes the Monte Carlo simulation on which the lifetime analysis is based. Nuclear states populated in low-energy binary heavy-ion reactions are considered, with decay time in the range of tens-to-hundreds femtoseconds, \textit{i.e.}, of the target-crossing time. The procedure consists of three major steps:
i) a preliminary Monte Carlo calculation to reconstruct the Total Kinetic Energy Loss (TKEL) distribution, for the population of a given nuclear state of the  projectile-like product (Sec. \ref{sec:4.1}),  ii) the simulation of the AGATA Doppler-shift corrected $\gamma$-ray spectrum, which is based on the projectile-like velocity calculated from the reconstructed TKEL (Sec. \ref{sec:4.2}), and iii) the two-dimensional $\chi^2$ minimization procedure, in lifetime-transition energy coordinates, based on the comparison between simulated and experimental $\gamma$-transition lineshapes (Sec. \ref{sec:4.3}). 


\subsection{Reconstruction of the initial velocity distribution}
\label{sec:4.1}

\begin{figure}[ht]
\centering
\resizebox{0.5\textwidth}{!}{\includegraphics{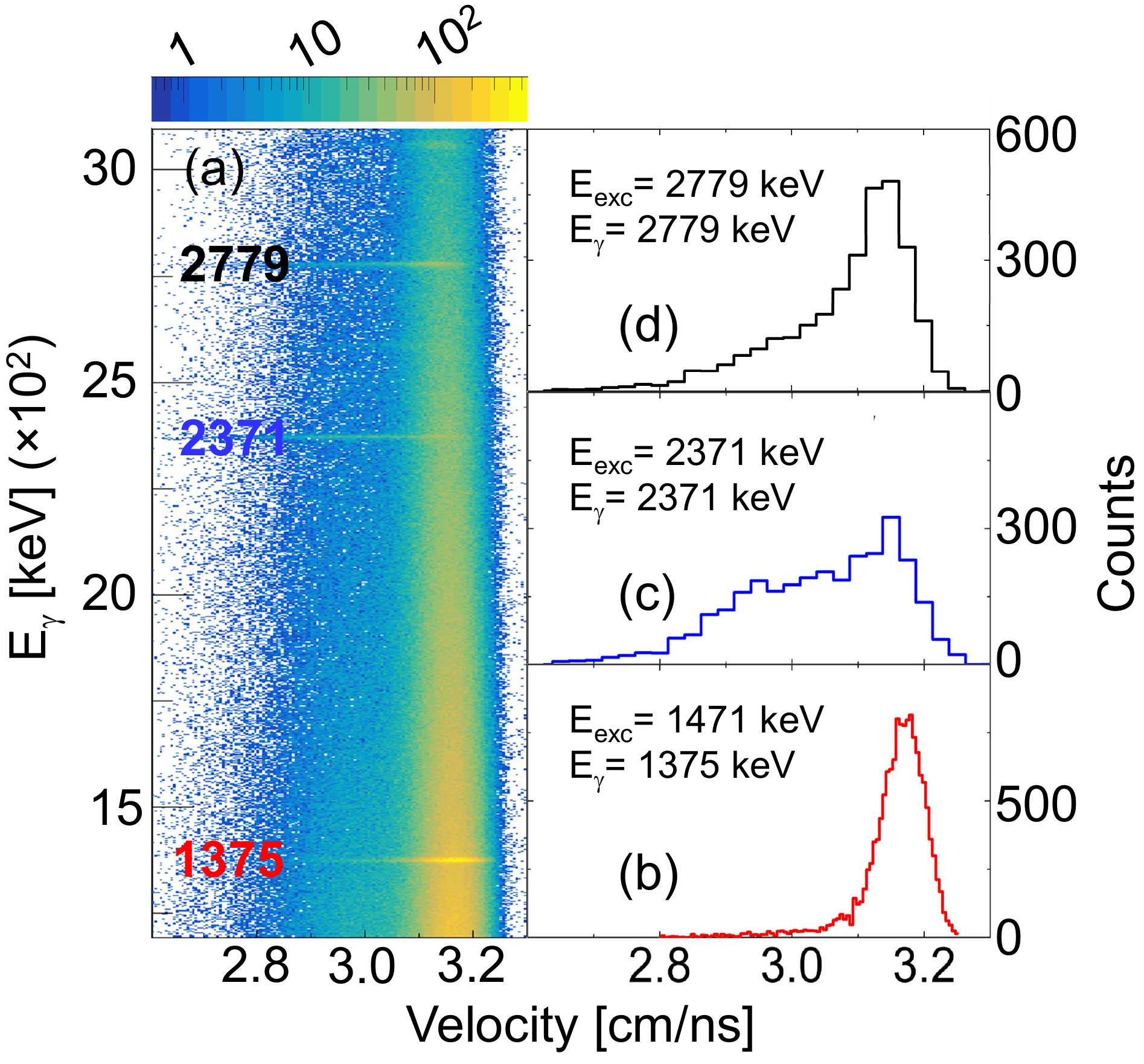}}
\caption{Two-dimensional plot of AGATA Doppler-shift corrected $\gamma$-ray energies vs. ions velocities, as measured in VAMOS++, in the case of $^{19}$O (a). Velocity distributions obtained by gating on the 1375- (b), 2371- (c) and 2779-keV (d) $\gamma$ rays depopulating the 1471-, 2371- and 2779-keV excited states of $^{19}$O. } 
\label{velocity_exp}
\end{figure}

The key point of the entire procedure is the determination of the velocity vector of the projectile-like product at the reaction instant, for a given excited state population. In the case of low-energy binary heavy-ion reactions, the velocity distribution of the reaction product includes contributions from both direct (quasi-elastic) and more dissipative processes, which lead to the appearance of broad structures at lower velocities \cite{Sch85,Kau61,Wil73,Zag14,Kar17,Ste18}. This is demonstrated in Fig. \ref{velocity_exp}, in the case of $^{19}$O. Panel (a) shows the matrix E$_{\gamma}$ vs. measured ion velocity for AGATA Doppler-shift corrected $\gamma$-rays. Transitions of energies 1375, 2371 and 2779 keV are clearly visible, depopulating excited states at 1471, 2371 and 2779 keV (see level scheme in Fig. \ref{19O}). Panels (b)-(d) show velocity distributions gated on each transition: only in the case of the 1/2$^+$ state at 1471 keV (b), a Gaussian-like velocity distribution is observed, while the other states display significant contributions at lower velocities from dissipative processes. 

\begin{figure*}[ht]
\centering
\resizebox{1.05\textwidth}{!}{\includegraphics{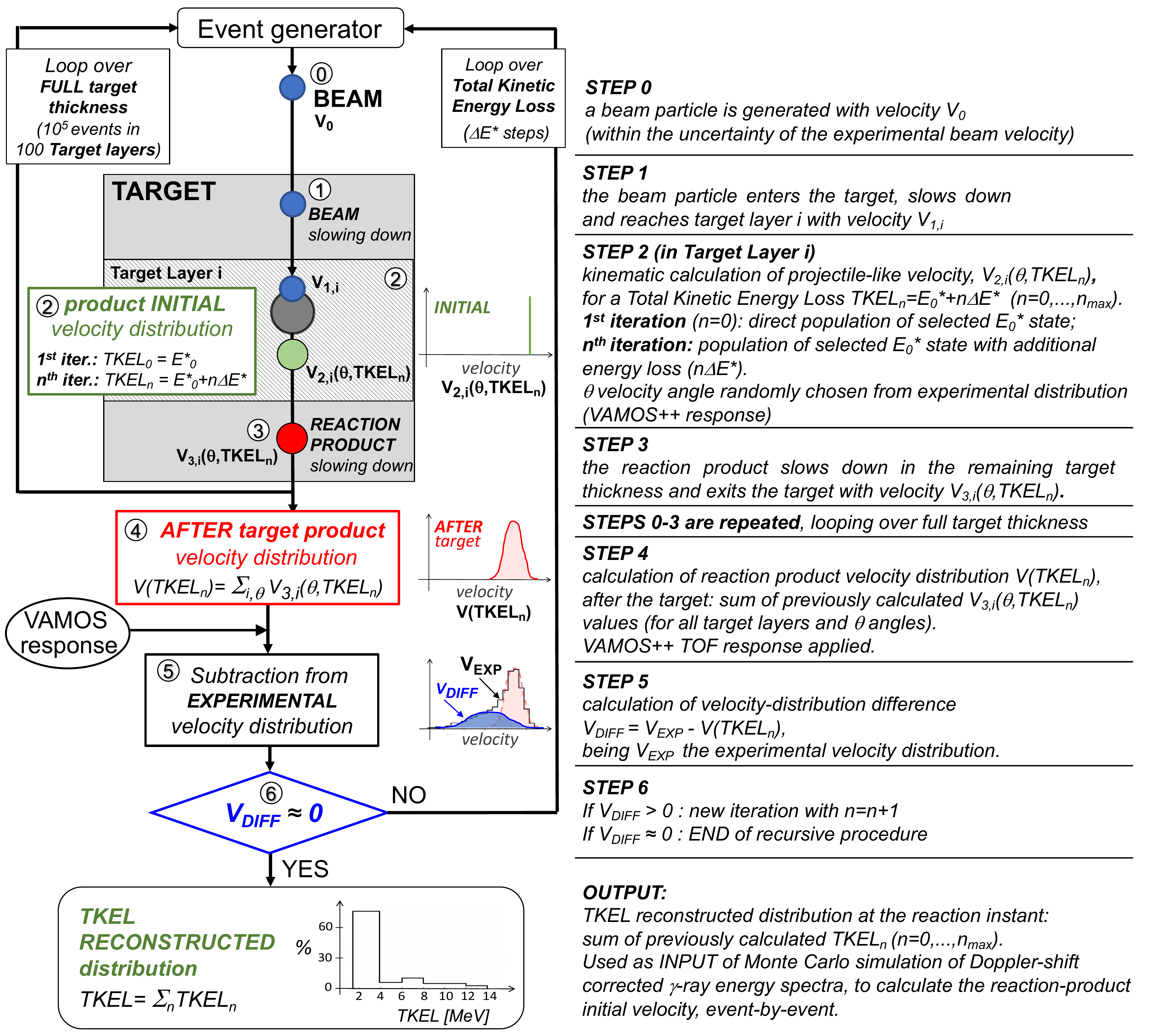}}
\caption{Left: Flow diagram of the Monte Carlo recursive procedure used to reconstruct the TKEL distribution (for a given projectile-like excited state E$^*_0$), on the basis of the measured velocity. Right: description of the steps of the procedure. } 
\label{FLOW_CHART}
\end{figure*}

Since dissipative contributions cannot be reliably calculated, a Monte Carlo procedure had to be developed to reconstruct the TKEL distribution, where TKEL is defined as the difference between the total kinetic energy before and after the collision. This calculation is performed prior to the simulation of the $\gamma$-ray emission (which is the subsequent step discussed in Sec. \ref{sec:4.2}) and it is based on a recursive subtraction, from the measured velocity distribution, of the velocity components associated with consecutive bins of TKEL. 

\begin{figure}[ht]
\centering
\resizebox{0.4\textwidth}{!}{\includegraphics{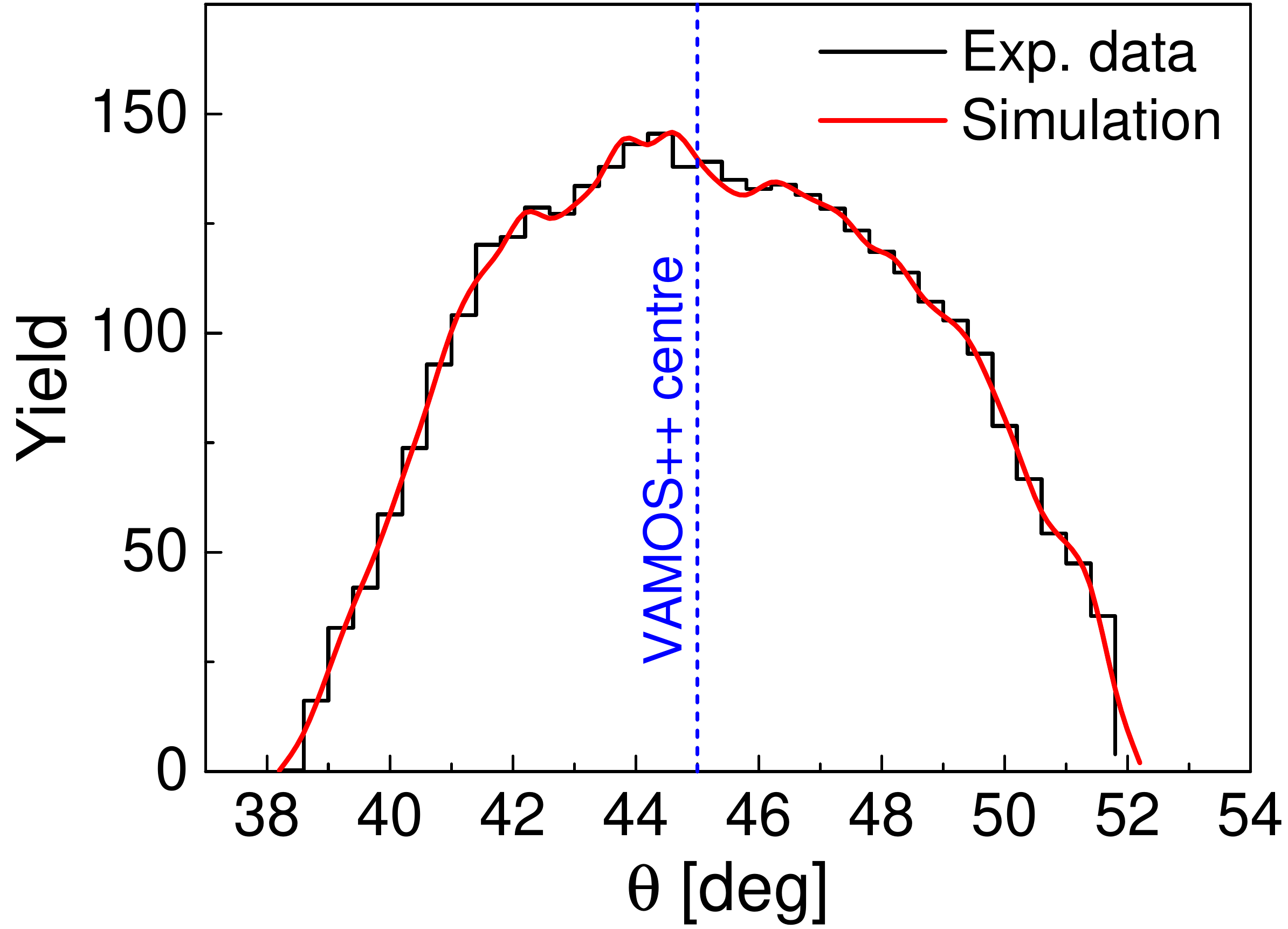}}
\caption{Comparison between the $\theta$ distribution of the $^{19}$O ions, associated with the de-excitation of the 2779-keV state, as measured in VAMOS++ (black histogram) and the corresponding Monte Carlo calculations (red). } 
\label{velocity_angle}
\end{figure}

As illustrated in the flow diagram of Fig. \ref{FLOW_CHART}, in the first iteration the velocity component associated with the direct population of the state of interest, at energy E$_0^*$, is calculated from the reaction kinematics.  The calculation is done looping over 100 target layers into which the target was divided, with the assumption that the reaction occurs at random over the full target thickness. In the case of the present reaction (with $\beta \sim$10$\%$), in each layer the beam-energy degradation is about 0.1 MeV, resulting in an energy loss of $\sim$10 MeV in the full target. The slowing down of the beam and projectile-like reaction product in the target is also considered. For each reaction event, the projectile-like product velocity direction is chosen based on the angular distribution measured in VAMOS++, which automatically takes care of the acceptance of the magnetic spectrometer \cite{Pul08} (see Fig. \ref{velocity_angle}). The final velocity distribution, after the target, is  corrected for the VAMOS++ ToF response (with a time uncertainty $\sigma$ = 1 ns), resulting in a smearing of the simulated velocity distribution. The final simulated velocity distribution is then subtracted from the experimental one, after proper normalization. 

In the next iteration, the second component of the projectile-like velocity distribution, associated with a TKEL increase by $\Delta$E$^*$, is calculated following a procedure similar to the first iteration, and the corresponding final velocity distribution, after the target, is also corrected for the VAMOS++ ToF response and subtracted from the remaining measured velocity distribution. In the present analysis, a number $n_{max}$ of 10 iterations was considered (with $\Delta$E$^*$ = 2 MeV), in order to fully reproduce the experimental velocity distribution, although a few iterations were found sufficient in all treated cases.

As a result of this preliminary Monte Carlo calculation, the TKEL distribution associated with the population of a given E$^*_0$ state of the  projectile-like product is reconstructed (see bottom of Fig. \ref{FLOW_CHART}). Such a TKEL distribution will be used in the main simulation of the AGATA Doppler-shift corrected $\gamma$-ray spectrum to calculate, event-by-event, the projectile-like product velocity vector, at the reaction instant, from the reaction kinematics (see Sec. \ref{sec:4.2}).

\begin{figure}[ht]
\centering
\resizebox{0.4\textwidth}{!}{\includegraphics{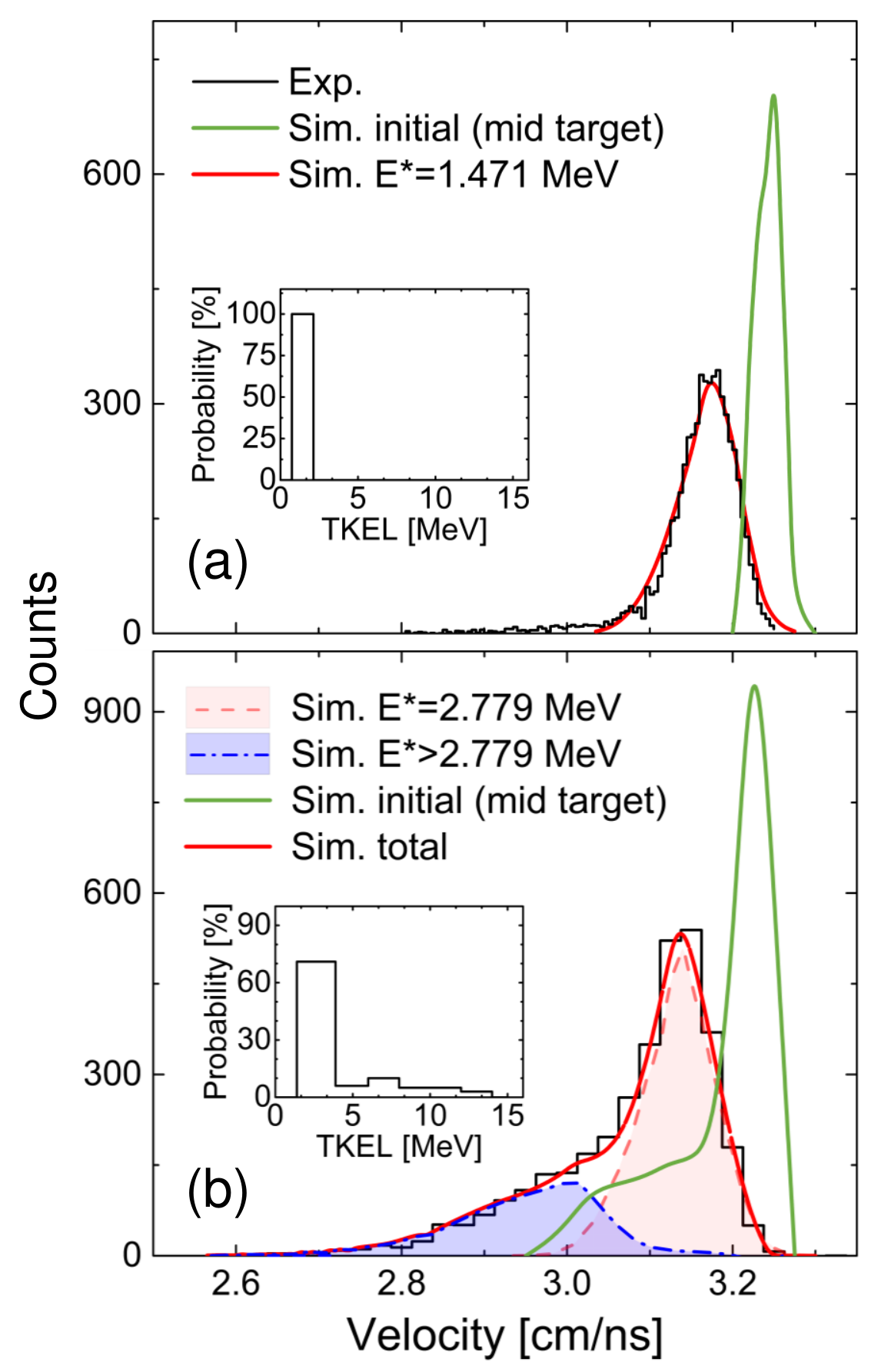}}
\caption{Panel (a): Measured (black histogram) and simulated (red solid line) velocity distributions for the de-excitation of the 1471-keV state of $^{19}$O. The unfolded initial velocity distribution, for reactions occurring at mid target, is shown by the green line. Panel (b): Same as in panel (a), for the de-excitation of the 2779-keV state of $^{19}$O. Red and blue distributions correspond to simulated direct (quasi-elastic) and dissipative components. The inset, in both panels, gives the reconstructed TKEL distribution for the corresponding state population (see Fig. \ref{FLOW_CHART} and text for details). } 
\label{velocity_sim}
\end{figure}

Figure \ref{velocity_sim} displays, as examples, the velocity distributions for the $^{19}$O product excited to the 1471-keV (a) and 2779-keV (b) states, and the corresponding simulated distributions. The measured (black histogram) and simulated (red line) velocity distributions have Gaussian-like shapes in the case of the 1471-keV state (panel (a)), which is characteristic for an exclusive direct population process. In contrast, the velocity distribution measured for the 2779-keV state (panel (b)) has a complex structure which is well described by a velocity profile with two separate contributions, associated with direct (red dashed) and dissipative (blue dashed) processes, respectively. In both panels, the solid green line displays the unfolded initial velocity distribution, for reactions occurring in a mid-target layer, while the inset gives the reconstructed TKEL distribution for the corresponding state population.

As mentioned above, an important ingredient of the simulation is the stopping-power parameterization, which was taken from Ziegler {\it{et al.}} \cite{Zie84,Ann88}. We evaluated the influence of this choice for our reaction by varying up to {20$\%$ the prescribed value, in the case of a uniquely determined reaction kinematics (i.e., the direct population of the 1471-keV state in $^{19}$O).  As shown in Fig. \ref{velocity_ziegler},  the simulated final velocity distribution (red band) reproduces the measured velocity profile within a 1$\sigma$ uncertainty for stopping power variation $<$5$\%$. We also compared the simulated velocity distributions extracted using the Ziegler {\it{et al.}} parametrization with different stopping-power laws, such as the one used by the code ATIMA \cite{Gei02}, which is usually considered for higher energies. Differences in energy losses were of the order of 2-3$\%$, resulting in negligible effects in the subsequent analysis.

\begin{figure}[ht]
\centering
\resizebox{0.45\textwidth}{!}{\includegraphics{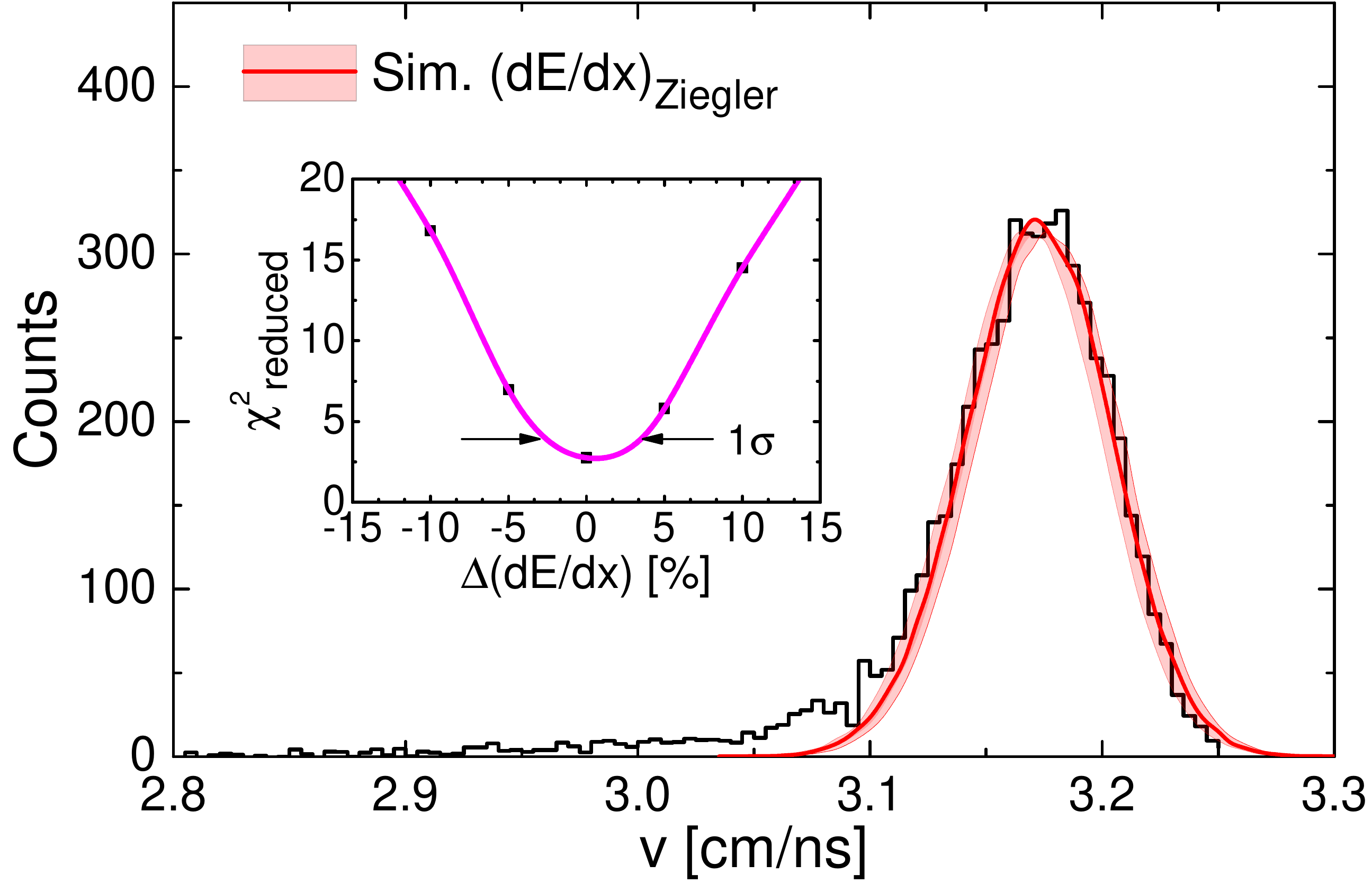}}
\caption{Measured velocity distribution for the de-excitation of the 1471-keV state in $^{19}$O. The red-shaded band is obtained by varying by 5$\%$ the stopping power parametrization of Ziegler {\it{et al.}} \cite{Zie84}. Inset: reduced $\chi^2$ curve for Monte Carlo simulations using different $\Delta$(dE/dx) variation (in $\%$) of the Ziegler stopping power parametrization, pointing to a 1$\sigma$ uncertainty  $<$5$\%$. } 
\label{velocity_ziegler} 
\end{figure}

\subsection{Simulation of AGATA Doppler-shift corrected $\gamma$-ray spectra }
\label{sec:4.2}

Simulation calculations of an AGATA Doppler-shift corrected $\gamma$-ray lineshape, for a transition depopulating a selected state, were performed in a two-step process. 

In the first step, the $\gamma$ events are prepared with a Monte Carlo procedure (with typical number of events of the order of 4$\times$10$^6$), following a flow diagram similar to the one presented in Fig. \ref{FLOW_CHART}. For each event, after the beam particle reaches the target layer in which the reaction occurs, the velocity of the projectile-like product, at the reaction instant, is obtained from a kinematic calculation assuming a TKEL value randomly chosen from the previously reconstructed TKEL distribution for the specific populated state (see Sec. \ref{sec:4.1}). The velocity direction is again randomly selected within the angular distribution measured in VAMOS++.  A $\gamma$ emission from the projectile-like product, slowing down in the remaining target thickness, is simulated for a given $\gamma$-transition energy and a decay time randomly chosen on the basis of the excited-state lifetime. 

In the second step, the $\gamma$-ray events generated by the Monte Carlo procedure discussed above are passed to the AGATA simulation package \cite{LabAGATA}, which provides, as an output, the $\gamma$-ray energy deposited in the AGATA crystals. In the calculations, the AGATA-simulation code considers the actual configuration of the array in the measurement ({\it{i.e.}}, 31 crystals with the corresponding geometry of the $^{18}$O (126 MeV) +  $^{181}$Ta experiment \cite{Cie20}).  Figure \ref{AGATA_sim} shows a comparison between the experimental and simulated $\gamma$-ray interaction positions projected on the x-y plane of AGATA, and associated with the first interaction point in the detector crystals. The strong similarity between the two distributions gives support to the quality of the AGATA-simulation code. 

\begin{figure}[ht]
\centering
\resizebox{0.47\textwidth}{!}{\includegraphics{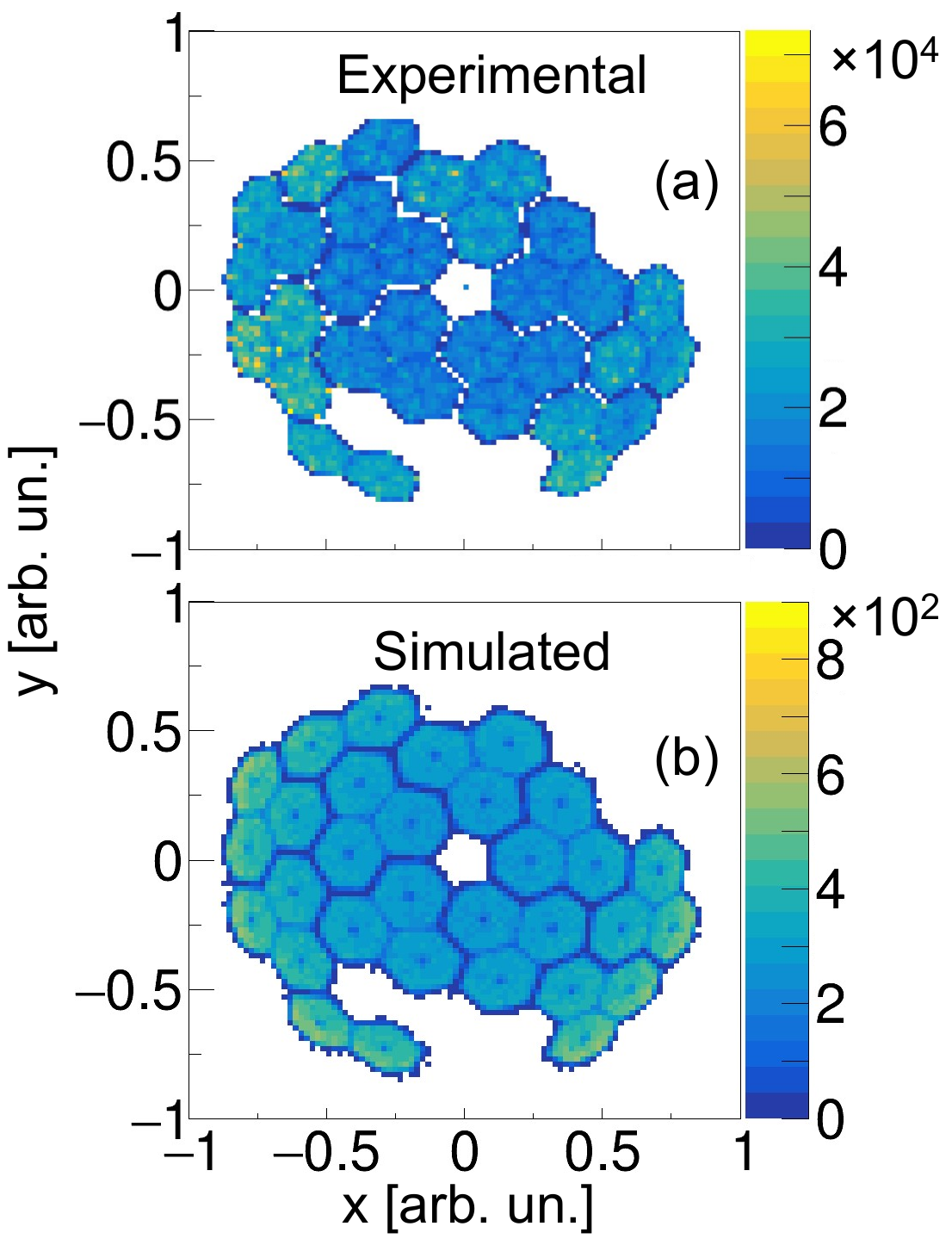}}
\caption{Experimental view of the x-y plane projection of the $\gamma$-ray interaction positions of the AGATA crystals (a) compared to  the simulated ones (b), in the case of the present $^{18}$O (126 MeV) +  $^{181}$Ta experiment \cite{Cie20}. The axes report the x and y positions of the first interaction point divided by the radius of the AGATA array.} 
\label{AGATA_sim}
\end{figure}

The simulated $\gamma$-ray data are subsequently analyzed with the AGATA OFT (Orsay Forward Tracking) algorithm \cite{Lop04}, following the same procedure applied to the experimental data. This allows to obtain the $\gamma$-ray energy and the relative angle between the $\gamma$-ray direction (reconstructed with the AGATA simulation package \cite{LabAGATA}) and the ion velocity vector, at the decay point (resulting from the Monte Carlo simulation procedure performed in the first step, as discussed above). The $\gamma$-ray Doppler-shift correction is then performed.  At this step, corrections are included to take into account the actual experimental energy resolution and the counting statistics of the AGATA detectors. 

\subsection{$\chi^2$ analysis of Doppler-broadened lineshapes for lifetimes determination}
\label{sec:4.3}

The evaluation of nuclear-state lifetimes in the time range of hundreds femtoseconds requires a detailed study of Doppler-broadened $\gamma$-ray lineshapes, as a function of the relative angle $\theta_{rel}$ between the moving-source and the emitted  $\gamma$-ray directions. Figure \ref{lifetime_DSAM} shows examples of simulated lineshapes for the 2.779-MeV $\gamma$ ray of $^{19}$O, over the full continuous-angle range $\theta_{rel}$ = 0-180$^{\circ}$. In the $^{18}$O+$^{181}$Ta reaction (sketched in panel (a)), such a transition de-excites the 2.779-MeV state populated by both direct and dissipative processes, as shown in Fig. \ref{velocity_sim} (b). In the simulations, three lifetime values are considered, {\it{i.e.}}, $\tau$ = 20, 100 and 2000 fs (panels (b), (c) and (d), respectively). In the short lifetime cases, a significant distortion of the overall lineshape is observed, which is at the basis of the nuclear-state lifetime evaluation. We underline that a lifetime-analysis procedure based on such a continuous-angle distributions, which is possible with $\gamma$-tracking arrays, acquires a significantly enhanced sensitivity, with respect to  experiments done with conventional $\gamma$-ray arrays with detectors placed at discrete angles, relative to the beam axis (see also Fig. \ref{AGATA_sensitivity}). This improvement was already pointed out by C. Stahl et al., \cite{Sta17} and C. Michelagnoli et al., \cite{Mic12} for the restricted cases of reactions in which  products have well-defined velocity vectors, such as Coulomb-excitation, transfer and fusion reactions.  In our work, we broaden the applicability of such a continuous-angle technique to reactions with complex structure of product velocity distribution, as in the case of low-energy binary, dissipative collisions. Crucial in this case is both the precise reconstruction of the emitted $\gamma$-ray direction by the tracking array, as well as the reaction-product-direction measurement by a magnetic spectrometer. 

\begin{figure}[htp]
\centering
\resizebox{0.4\textwidth}{!}{\includegraphics{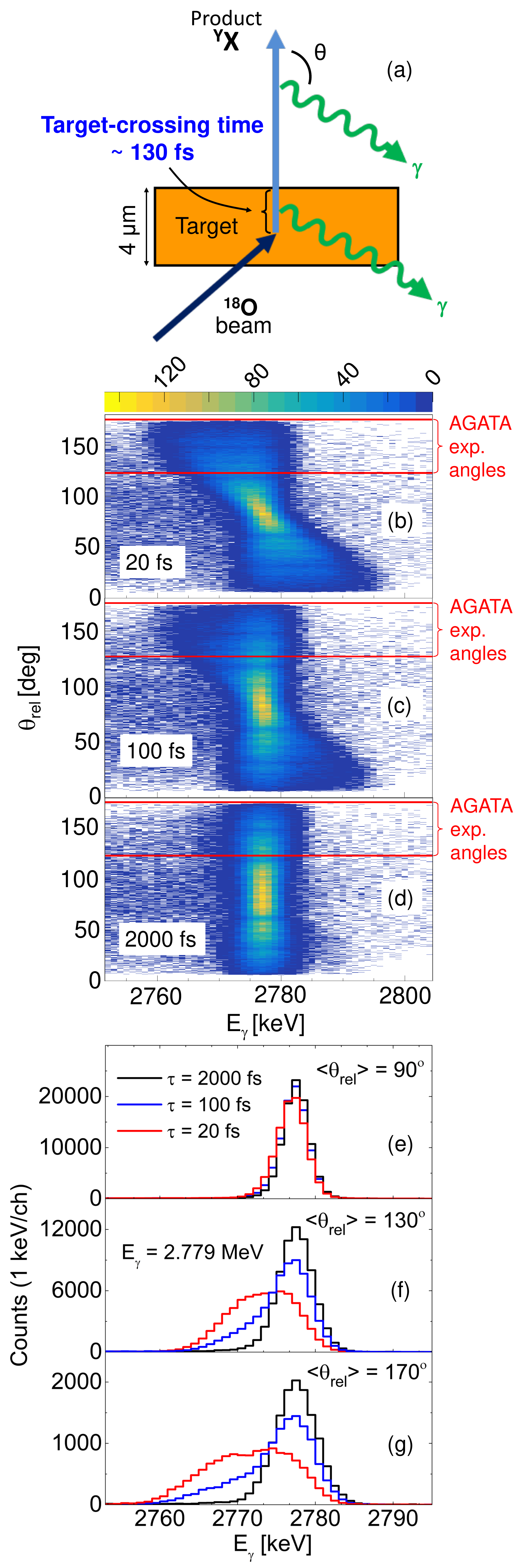}}
\caption{Panel (a): illustration of the interaction process of an $^{18}$O 126-MeV beam in a  4-$\mu$m thick  $^{181}$Ta target, resulting in a $^Y$X reaction product which de-excites by $\gamma$-ray emission. $\theta$ is the angle between the reaction product velocity vector (as measured in the VAMOS++ spectrometer, for example) and the emitted $\gamma$-ray direction. 
Panels (b)-(d): simulated two-dimensional (E$_{\gamma}$, $\theta$) lineshapes, for a 2.779-MeV $\gamma$ ray from a 2.779-MeV state of $^{19}$O, assuming lifetime values of 20 (b), 100 (c) and 2000 (d) fs. Horizontal lines give the AGATA angular coverage of the GANIL experiment. Panels (e)-(g): projections on the $\gamma$-ray energy axis for $\theta$ = 90$^{\circ}\pm$10$^{\circ}$ (c),  130$^{\circ}\pm$10$^{\circ}$ (d) and 170$^{\circ}\pm$10$^{\circ}$ (e).} 
\label{lifetime_DSAM}
\end{figure}

A closer view of the Doppler-broadened lineshapes is given by the projections of the simulated matrices on the $\gamma$-ray-energy axis for $\theta_{rel}$ =  130$^{\circ}\pm$10$^{\circ}$ and 170$^{\circ}\pm$10$^{\circ}$ (panel (f) and (g), respectively), which correspond to the angular coverage of the AGATA array in the GANIL experiment. It is seen that, for lifetime values much longer than the target-crossing time ({\it{i.e.}}, $>>$130 fs), a symmetric lineshape, with the same centroid energy, is observed at all angles (black histograms) after a Doppler-shift correction based on the ion final velocity ({\it{i.e.}}, after the target). A considerable lineshape distortion is instead observed for shorter lifetimes, what can be used to obtain a precise estimate of $\tau$ for lifetimes of the order of the target-crossing time (blue lines). The lower limit for lifetime determination is found to be of the order of tens of fs, at which the $\gamma$ line becomes broad and featureless, and significantly shifted in energy (red histograms). 
As discussed later, crucial for the analysis is also the precise $\gamma$-ray energy determination provided by the 90$^{\circ}$ detectors (panel (c)), which are not affected by the Doppler shift. Altogether, this clearly indicates that the best conditions for precise lifetime determination will be reached by a tracking array with an extended angular coverage, as it is foreseen for AGATA in the coming future \cite{Kor20}.

Coming now to the details of the lifetime analysis procedure, here developed, the technique relies on a two-dimensional $\chi^2$ minimization, in lifetime and transition energy coordinates (E$_{\gamma}$, $\tau$). The two-dimensional $\chi^2$ surface is expected to show  a minimum corresponding to the optimal state lifetime and transition energy.  Figures \ref{tau_17O} and \ref{tau_19O} show examples of lifetime analyses for $^{17}$O and $^{19}$O states for which $\tau$ values of the order of 100 fs are reported in literature \cite{NNDC}, {\it{i.e.}}, well within the sensitivity range of the present technique. In the first case, the 3055-keV,  1/2$^-_1$  state in $^{17}$O, which is depopulated by a 2184-keV $\gamma$ ray, is considered (see level scheme in Fig. \ref{17O}). Simulated and experimental $\gamma$-ray spectra are compared in three ranges of the relative angle $\theta_{rel}$: 120$^{\circ}$-140$^{\circ}$ , 140$^{\circ}$-160$^{\circ}$  and 160$^{\circ}$-180$^{\circ}$, as shown in panels (a), (b) and (c) of Fig. \ref{tau_17O}. The corresponding two-dimensional $\chi^2$ lifetime-energy surface is reported in panel (d). A well-defined minimum (marked with a white cross) is visible at $\tau$=159$^{+40}_{-30}$ fs and E$_{\gamma}$= 2184.3$^{+0.3}_{-0.2}$ keV, in agreement, within uncertainty, with the literature values of $\tau$=120$^{+80}_{-60}$ fs and E$_{\gamma}$=2184.44(9) keV \cite{NNDC}. The uncertainties are obtained by considering the 1$\sigma$ region around the optimum value, as indicated with a white contour in panel (d). The red shaded bands in panels (a)-(c) are the results of the lineshape simulations performed by varying E$_{\gamma}$ and $\tau$ within the 1$\sigma$ region around the $\chi^2$ minimum. 

A similar analysis is reported in Fig. \ref{tau_19O} for the 2779-keV, 7/2$^+$ state in $^{19}$O, deexcited by a 2779-keV transition (see level scheme in Fig. \ref{19O}). Also in this case, a well-defined minimum is found in the $\chi^2$ map, located at $\tau$ = 140$^{+50}_{-40}$ fs and E$_{\gamma}$ = 2779.0$^{+1.0}_{-0.8}$ keV, in line with previous works (i.e., $\tau$ = 70(26) fs \cite{Brou71} and $\tau$ = 117(26) fs \cite{Hib71}). 

An additional example of lifetime analysis, within the sensitivity range of the present technique, is reported in Ref. \cite{Cie20}, for the second 2$^+$ state in $^{20}$O, located at 4070 keV excitation energy, with a lifetime $\tau$ = 150$^{+80}_{-30}$ fs (see later discussion in connection with Fig. \ref{tau_sensitivity}).

\begin{figure*}[htp]
\centering
\resizebox{1.0\textwidth}{!}{\includegraphics{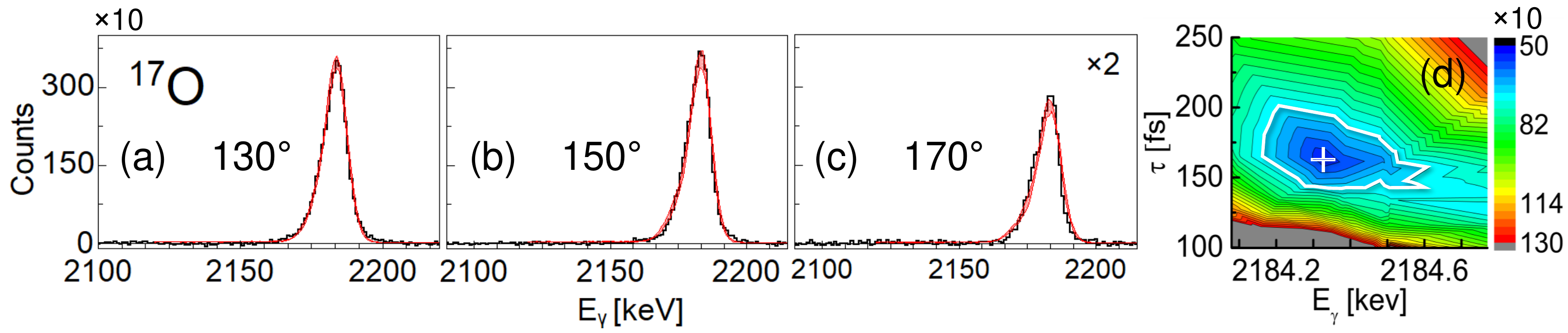}}
\caption{Doppler-shift corrected AGATA $\gamma$-ray energy spectra (black) and simulated ones (red) in the region of the 2184-keV line of  $^{17}$O, for the angular ranges of 130$^{\circ}\pm 10^{\circ}$ (a), 150$^{\circ}\pm 10^{\circ}$ (b), and 170$^{\circ}\pm 10^{\circ}$ (c). Panel (d): corresponding two-dimensional $\chi^2$ lifetime-energy surface, with the white cross and white contour line indicating the minimum and the 1$\sigma$  uncertainty region.  The red shaded bands in panel (a)-(c) are the results of the lineshape simulations performed by varying E$_{\gamma}$ and $\tau$ within the 1$\sigma$ region, {\it{i.e.}}, (E$_{\gamma}$, $\tau$) = (2184.3$^{+0.3}_{-0.2}$,159$^{+40}_{-30}$ fs ). } 
\label{tau_17O}
\end{figure*}

\begin{figure*}[ht]
\centering
\resizebox{1.0\textwidth}{!}{\includegraphics{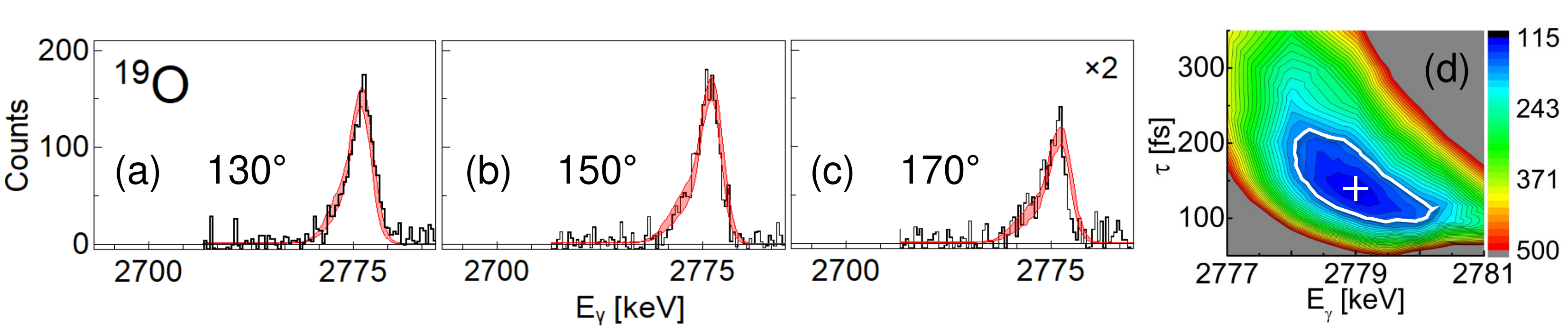}}
\caption{Doppler-shift corrected AGATA $\gamma$-ray energy spectra (black) and simulated ones (red) in the region of the 2779-keV line of $^{19}$O, for the three angular ranges of 130$^{\circ}\pm 10^{\circ}$ (a), 150$^{\circ}\pm 10^{\circ}$ (b), and 170$^{\circ}\pm 10^{\circ}$ (c). Panel (d): corresponding two-dimensional $\chi^2$ lifetime-energy surface, with the white cross and white contour line indicating the minimum and the 1$\sigma$  uncertainty region.  The red shaded bands in panel (a)-(c) are the results of the lineshape simulations performed by varying E$_{\gamma}$ and $\tau$ within the 1$\sigma$ region, {\it{i.e.}}, (E$_{\gamma}$, $\tau$) = (2779.0$^{+1.0}_{-0.8}$ keV,140$^{+50}_{-40}$ fs).} 
\label{tau_19O}
\end{figure*}


\begin{figure*}[ht]
\centering
\resizebox{1.0\textwidth}{!}{\includegraphics{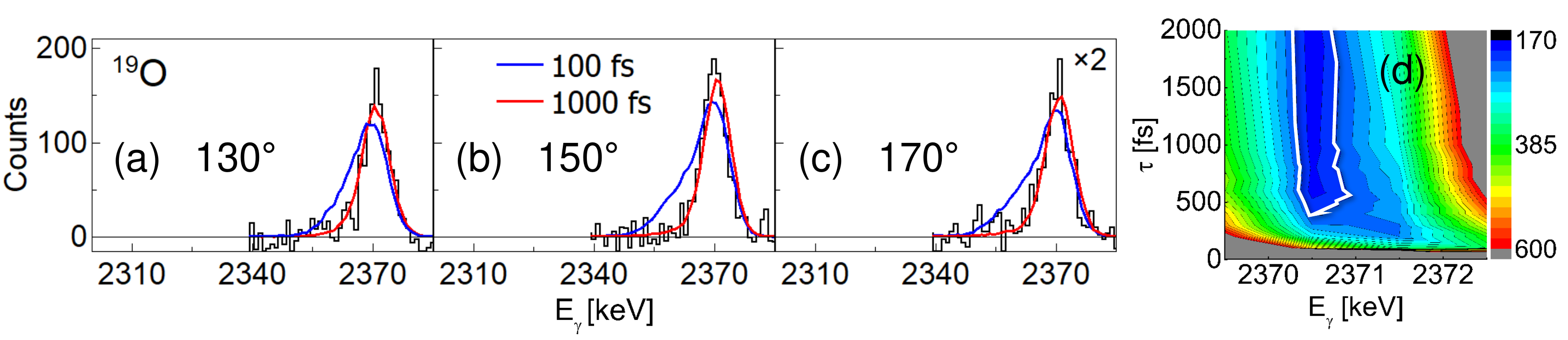}}
\caption{Doppler-shift corrected AGATA $\gamma$-ray energy spectra (black) and simulated ones for lifetime values of 100 fs (blue) and 1000 fs (red), in the region of the 2371-keV line of  $^{19}$O, for the three angular ranges of 130$^{\circ} \pm$10$^{\circ}$  (a), 150$^{\circ} \pm$10$^{\circ}$  (b), and 170$^{\circ} \pm$10$^{\circ}$ (c). Panel (d): corresponding two-dimensional $\chi^2$ surface in (E$_{\gamma}$, $\tau$) coordinates, with the white contour line delimiting the 1$\sigma$  uncertainty region (with a lower limit $\tau >$ 400 fs at E$_{\gamma}$ = 2370.6$^{+0.5}_{-0.3}$ keV).} 
\label{tau_19O_long}
\end{figure*}

\begin{figure*}[ht]
\centering
\resizebox{1.0\textwidth}{!}{\includegraphics{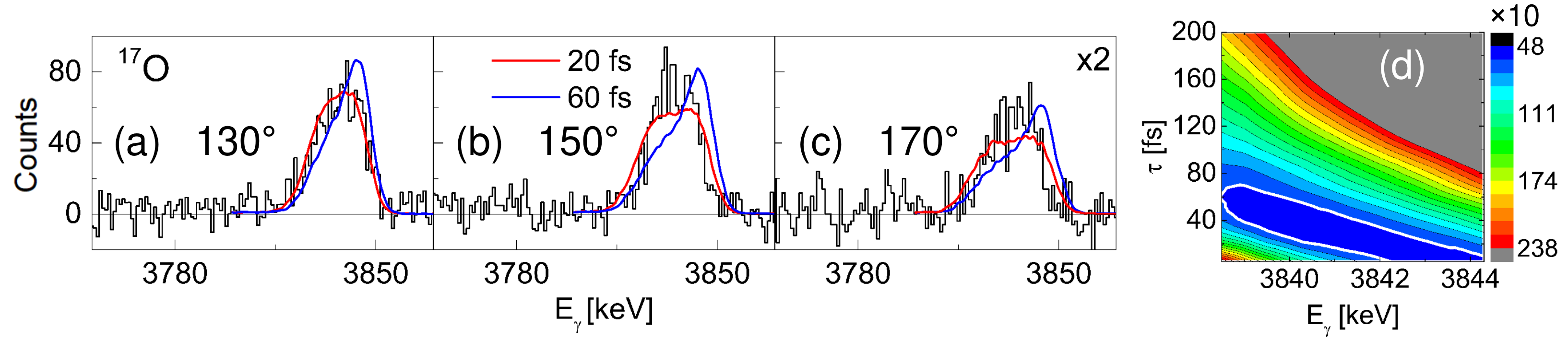}}
\caption{Doppler-shift corrected AGATA $\gamma$-ray energy spectra (black) and simulated ones with lifetime values of 20 fs (red) and 60 fs (blue), in the region of the 3842-keV line of  $^{17}$O, for the three angular ranges of 130$^{\circ} \pm$10$^{\circ}$  (a), 150$^{\circ} \pm$10$^{\circ}$  (b), and 170$^{\circ} \pm$10$^{\circ}$ (c). Panel (d): corresponding two-dimensional $\chi^2$ surface in (E$_{\gamma}$, $\tau$) coordinates, with the white contour line delimiting the 1$\sigma$ uncertainty region. } 
\label{tau_17O_short}
\end{figure*}

\subsection{Sensitivity limits of the lifetime analysis technique}
\label{sec:4.4}
As discussed in the previous section, the time range accessible by the present lifetime analysis technique is dictated by the target-crossing time T$_{cross}$ of the reaction product (which is about 130 fs for the $^{18}$O+$^{181}$Ta reaction). The simulation showed that this range spans between $\sim$0.2T$_{cross}$ and  $\sim$4T$_{cross}$. Consequently, for lifetime values a few times longer or shorter than T$_{cross}$,  the here-proposed two-dimensional $\chi^2$ minimization procedure, in lifetime-transition energy coordinates (E$_{\gamma}$, $\tau$), cannot provide a well localized minimum. Rather, a valley extending towards infinitely long lifetimes or reaching the zero value will be obtained.

To illustrate this aspect of sensitivity limit, we consider in Figure \ref{tau_19O_long} the case of the long-lived 2371-keV state in $^{19}$O, for which the lower limit $\tau >$ 3.5 ps is reported in literature \cite{Hib71}. Panels (a), (b) and (c) display the Doppler-shift corrected 2371-keV $\gamma$ ray (deexciting the state), as measured in AGATA in the three angular ranges of 120$^{\circ}$-140$^{\circ}$, 140$^{\circ}$-160$^{\circ}$ and 160$^{\circ}$-180$^{\circ}$, respectively. In all cases, a symmetric Gaussian lineshape is observed, as expected for decays occurring outside the target, at times significantly larger than the target-crossing time. Simulated lineshapes corresponding to $\tau$ = 100 and 1000 fs are also shown in blue and red, respectively, for comparison. No minimum is obtained in the $\chi^2$ map (see Fig. \ref{tau_19O_long}(d)), but a valley is observed, extending from $\tau >$ 400 fs, at the $\gamma$-transition energy of 2370.6$^{+0.5}_{-0.3}$ keV, which agrees well with the literature value. 

Figure \ref{tau_17O_short} shows the case of the short-lived 3842-keV state in $^{17}$O, for which the upper limit $\tau<$ 26 fs is reported in literature \cite{Ale64}. The Doppler-shift corrected 3842-keV $\gamma$ ray (deexciting the state) is shown in panels (a), (b) and (c), as measured in AGATA in the three angular ranges of 120$^{\circ}$-140$^{\circ}$, 140$^{\circ}$-160$^{\circ}$ and 160$^{\circ}$-180$^{\circ}$, respectively. In all cases, a broad-peak structure is observed around 3835 keV, which is consistent with a $\gamma$ emission, inside the target, from a very short-lived state. Simulated lineshapes corresponding to $\tau$= 20 and 60 fs are also shown in red and blue, respectively, for comparison. Also in this case, no well-defined minimum is found in the $\chi^2$ map (see Fig. \ref{tau_17O_short}(d)), but a valley is seen, extending from 70 fs down to 0 fs, with a strong dependence on the $\gamma$-ray transition energy. A lifetime $\tau$= 20$^{+20}_{-20}$ fs is obtained if the $\gamma$-transition energy is taken to be 3842.3(4) keV, as reported in literature \cite{NNDC}. This shows the impact of a precise $\gamma$-ray energy determination, which could be best accomplished when the tracking array extends to 90$^{\circ}$. 

\subsection{Relevance of the AGATA tracking array performances}
\label{sec:4.5}

The quality of the results of the newly developed lifetime analysis technique depends strongly on the Doppler-shift correction capabilities of the experimental setup. In the case of $\gamma$-ray tracking arrays, such as AGATA, the interaction point is identified with unprecedented precision, with respect to conventional HPGe detectors \cite{Akk12,Cle17,Kor20}. In a standard configuration of AGATA (at 23.5 cm from the target center), the angular resolution is around 1$^{\circ}$, as a result of the combined use of Pulse Shape Analysis and tracking algorithms. With the use of a magnetic spectrometer, which also offers a resolution of 1$^{\circ}$ for the angle detection of the reaction products (as in the case of VAMOS++), the angle between the fragment velocity at the de-excitation point and the $\gamma$-ray direction can be determined with an accuracy of about 1.5$^{\circ}$. Such a precision is crucial, together with an accurate measurement of the ion velocity, to perform a Doppler-shift correction which allows for a detailed study of the $\gamma$-ray  lineshape, as discussed in this paper.

\begin{figure}[htp]
\centering
\resizebox{0.4\textwidth}{!}{\includegraphics{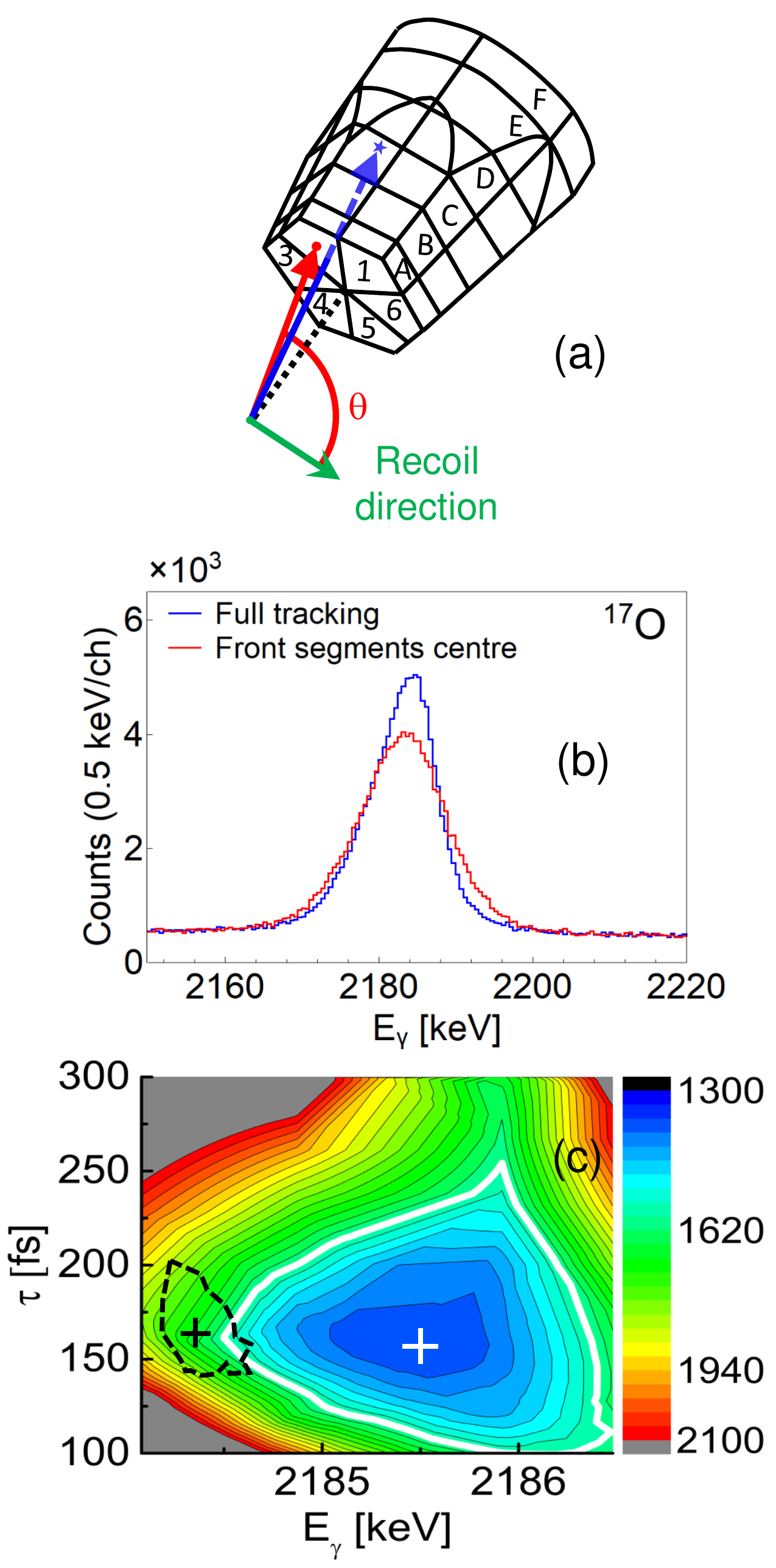}}
\caption{Panel (a):  interaction of a $\gamma$-ray, emitted from a recoiling reaction fragment, in a segmented AGATA germanium detector. The  real $\gamma$-interaction point  is marked by a blue star,  the front-segment center by a red point. Panel (b): AGATA Doppler-shift corrected 2184-keV, 1/2$^-_1$ $\rightarrow$1/2$^+_1$, $\gamma$ transition of $^{17}$O, obtained by applying a Doppler-shift correction based on the full tracking procedure (blue) or considering the front-segment center of the AGATA detector (red), as $\gamma$-ray interaction point. Panel (c): lifetime-energy $\chi^2$ minimization surface for the 2184-keV transition, obtained by considering the front-segment centers of the AGATA detectors as $\gamma$-ray interaction point. The white cross and contour line indicate the minimum and the 1$\sigma$ uncertainty region. The dashed contour, in black, delimits the 1$\sigma$ uncertainty region obtained with a full-tracking analysis of the AGATA data, as already reported in Fig. \ref{tau_17O}(d).} 
\label{AGATA_sensitivity}
\end{figure}

Figure \ref{AGATA_sensitivity}(b) gives, as an example, the lineshape of the 2184-keV $\gamma$ ray of $^{17}$O obtained by determining the $\gamma$-ray interaction points using the full AGATA tracking procedure (blue histogram), or by considering the front-segment centers (red histogram), as it is done with conventional HPGe detectors (for the determination of the interaction points, see Fig. \ref{AGATA_sensitivity}(a)). In the latter case, the less precise Doppler-shift correction is found to limit significantly the lineshape sensitivity to the lifetime and $\gamma$-ray energy determination. As shown in Fig. \ref{AGATA_sensitivity}(c), a shallower minimum is obtained in the lifetime-transition energy $\chi^2$ surface with respect to Fig. \ref{tau_17O}(d), leading to much larger uncertainties in the final E$_{\gamma}$ and $\tau$ values. In the specific case of the 2184-keV $\gamma$ ray of $^{17}$O, the 1$\sigma$ region is $\sim$5  times more extended in E$_{\gamma}$ and $\sim$2.3 times wider in $\tau$. Moreover, the $\chi^2$ value at the minimum of the two-dimensional map is $\sim$2.5 times larger than in the AGATA analysis performed with full tracking.  A similar behavior of the $\chi^2$ map was observed (when the front-segment centers were considered) in the lifetime analyses of the 7/2$^+$ and second 2$^+$ states in $^{19}$O and $^{20}$O \cite{Cie20}, respectively. Figure \ref{tau_sensitivity} displays the results of the lineshape analysis in the case of Doppler-shift corrections based on full tracking (square symbols with error bars) and on interaction positions taken as front-segment-center (contour areas), in the (E$_{\gamma}$,$\tau$) coordinates. For the two test cases of $^{17}$O and $^{19}$O, the NNDC adopted values \cite{NNDC} are denote by circles with error bars. It is seen that the lifetime and energy determination is quite accurate if the tracking procedure is applied, while the analysis based on the front-segment-center interaction positions suffers from large uncertainties in both lifetime and $\gamma$-energy coordinates. We note that these uncertainties would be much larger in the case of conventional HpGe arrays, where  individual crystals are typically bigger than AGATA segments.

\begin{figure}[ht]
\centering
\resizebox{0.48\textwidth}{!}{\includegraphics{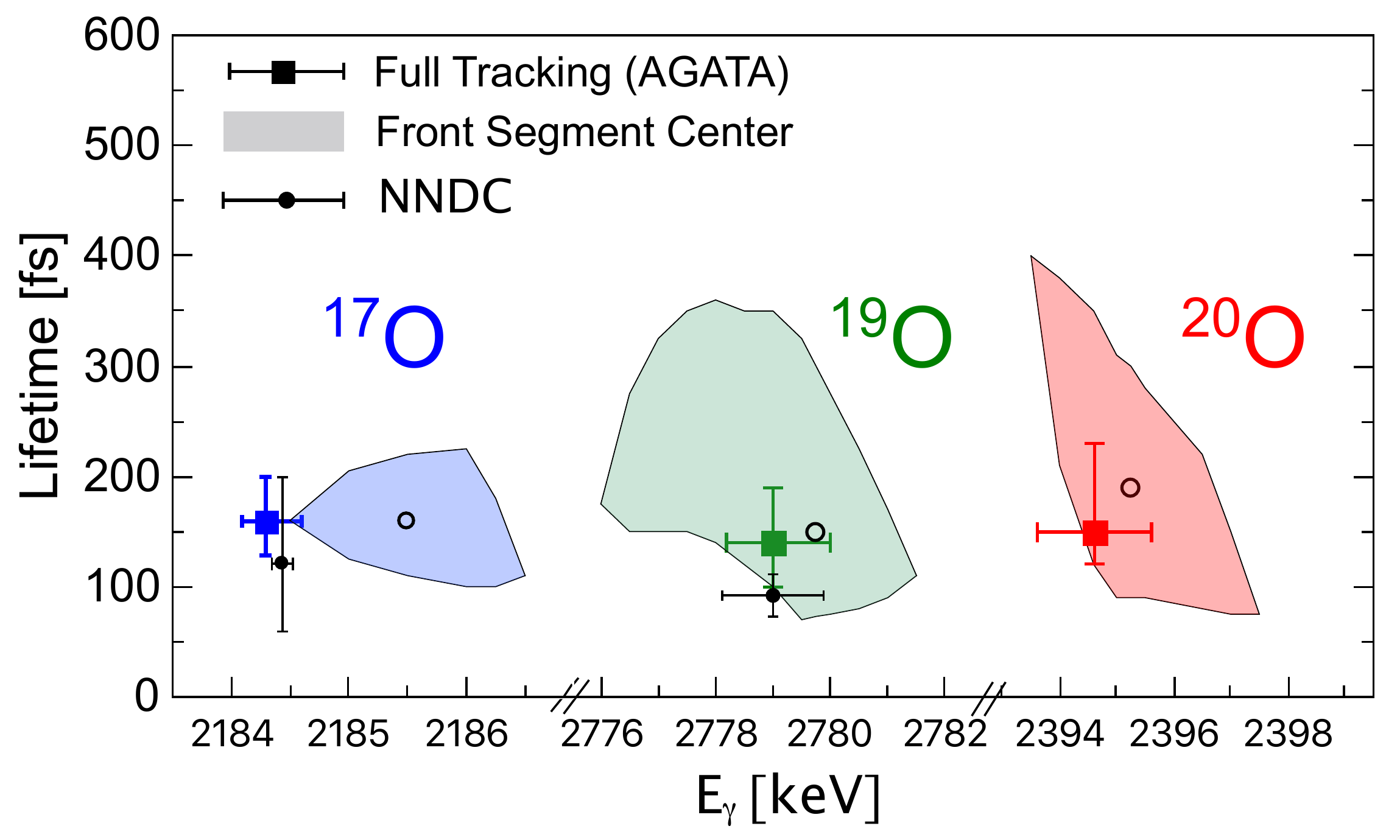}}
\caption{Lifetime values obtained with the present Monte Carlo technique for the 3055-keV, 1/2$^{-}$ state of $^{17}$O (blue), the 2779-keV, 7/2$^+$ state of $^{19}$O (green),  and the 4070-keV, second 2$^+$ state of $^{20}$O (red), considering the $\gamma$-interaction points extracted with the AGATA full-tracking procedure (square symbols) and the front-segment centers of the AGATA detectors (contour areas with open circles indicating the best values). For the two test cases of $^{17}$O and $^{19}$O, the NNDC adopted values \cite{NNDC} are given by circles symbols with error bars.} 
\label{tau_sensitivity}
\end{figure}

\section{Conclusions}
\label{sec:conclusion}

A novel Monte Carlo technique has been developed to determine nuclear-state lifetimes of the order of tens-to-hundreds femtoseconds (\textit{i.e.}, the target-crossing time), by accurate analysis of Doppler-broadened $\gamma$-ray lineshapes, in low-energy heavy-ion binary reactions. These reaction processes are characterized by large energy dissipation, leading to complex velocity distributions which do not allow to apply standard lineshape analysis methods. Our procedure makes use of the reaction-product velocity distribution, as measured by a magnetic spectrometer, to reconstruct, on event-by-event basis, the velocity at the decay instant. The latter is then used in the $\gamma$-ray Doppler correction calculations.

In the present paper, the method is discussed in connection with the analysis of an experiment performed at GANIL with the AGATA+VAMOS+PARIS setup, aiming at the study of excited states lifetimes in neutron-rich O, C, and N nuclei \cite{Cie20}. It is demonstrated that the combined use of a magnetic spectrometer and a $\gamma$-tracking array (with  few millimeter interaction-point position resolution) becomes essential  for the detailed analysis of the $\gamma$-ray lineshapes, resulting in state lifetime determinations. The method will significantly gain in precision when tracking arrays will reach a large angular coverage. The present work  clearly shows, as well, that tracking arrays are unique for precision $\gamma$-spectroscopy studies in low-energy reaction regimes,  in addition to their powerful application in collisions at relativistic energies, as demonstrated in earlier works \cite{Mor18,Kor20,Avi20,Bra15,Pod16,Bos19,Gad14,Str14}.

The new approach discussed in this work is expected to become an important tool for investigating exotic neutron-rich nuclei produced with intense ISOL-type beams in low-energy heavy-ion binary collisions: it will allow to obtain information on electromagnetic observables which can be used to test the quality of first-principles nuclear structure calculations, complementing common benchmarks based on nuclear-state energies.
 
\bigskip

This work was supported by the Italian Istituto Nazionale di Fisica Nucleare, by the Polish National Science
Centre under Contracts No. 2014/14/M/ST2/00738, No. 2013/08/M/ST2/00257, and No. 2016/22/M/ST2/00269,
and by RSF Grant No. 19-42-02014 and by the U.S. Department of Energy, Office of Science, Office of Nuclear
Physics, under contract number DE-AC02-06CH11357. This project has received funding from the Turkish Scientific and Research Council (Project No. 115F103) and from the European Union Horizon 2020 Research and Innovation Program under Grant Agreement No. 654002. 




\end{document}